\documentclass[reprint,amsmath,amssymb,aps,]{revtex4-2}
\UseRawInputEncoding
\usepackage{graphicx}
\usepackage{dcolumn}
\usepackage{bm}

\begin{document}

\preprint{APS/123-QED}

\title{Ultrafast photocurrents in MoSe$_2$ probed by terahertz spectroscopy}

\author{Denis Yagodkin$^1$}
\email{iagodkin@zedat.fu-berlin.de}

\author{Lukas Nadvornik$^{2,3}$}%
\thanks{The first two authors contributed equally}
\author{Oliver Gueckstock$^{1,2}$}
\author{Cornelius Gahl$^1$}
\author{Tobias Kampfrath$^{1,2}$}
\author{Kirill I. Bolotin$^1$}
\email{kirill.bolotin@fu-berlin.de}
\affiliation{$^1$Department of Physics, Freie Universität Berlin}%
\affiliation{$^2$Department of Physical Chemistry, Fritz Haber Institute of the Max Planck Society}
\affiliation{$^3$Faculty of Mathematics and Physics, Charles University}
\date{\today}

\begin{abstract}
We use the terahertz (THz) emission spectroscopy to study femtosecond photocurrent dynamics in the prototypical 2D semiconductor, transition metal dichalcogenide MoSe$_2$. We identify several distinct mechanisms producing THz radiation in response to an ultrashort ($30\,$fs) optical excitation in a bilayer (BL) and a multilayer (ML) sample. In the ML, the THz radiation is generated at a picosecond timescale by out-of-plane currents due to the drift of photoexcited charge carriers in the surface electric field. The BL emission is generated by an in-plane shift current. Finally, we observe oscillations at about $23\,$THz in the emission from the BL sample. We attribute the oscillations to quantum beats between two excitonic states with energetic separation of $\sim100\,$meV. 
\end{abstract}

\maketitle

Photoinduced carrier dynamics in semiconducting transition metal dichalcogenides (TMDCs) such as MoSe$_2$, MoS$_2$, or WSe$_2$ have been actively investigated recently \cite{near_unity_yeald_MoS2,dynamics_Ex_WSe2,Tr_Ex_MoSe2_and_WSe2}. The dynamics is especially interesting as optical excitations in TMDCs are dominated by excitons, strongly bound electron/hole complexes \cite{absorption_spectrum_TMDCs,Klots_Newaz_Wang_Prasai_Krzyzanowska_Lin_Caudel_Ghimire_Yan_Ivanov}. Weak screening of excitons leads to large binding energies up to hundreds of meV, making them stable at room temperature \cite{Chernikov_Berkelbach_Hill_Rigosi_Li_Aslan_Reichman_Hybertsen_Heinz_2014,Klots_Weintrub_Prasai_Kidd_Varga_Velizhanin_Bolotin_2018,Greben_Arora_Harats_Bolotin_2020,Poellmann_Steinleitner_Leierseder_Nagler_Plechinger_Porer_Bratschitsch_Schller_Korn_Huber_2015}. Rich dynamics have been shown to result from the interactions between free carriers, neutral excitons, and charged excitons (three-particle complexes) \cite{Singh_Moody_Tran_Scott_Overbeck_B,Dynamics_trions_free_carriers,Yu_Yu_Xu_Barrette_Gundogdu_Cao_2016}. In addition, monolayer TMDCs are two-dimensional materials which can be stacked into heterostructures  \cite{Novoselov_Mishchenko_Carvalho_Castro}. The dynamics of interlayer excitons (hole in one layer and electron in another) in such systems  can be controlled by the choice of materials \cite{Inter_Ex_MoS2WS2,Hanbicki_Chuang_Rosenberger_Hellberg_Sivaram_McCreary_Mazin_Jonker_2018}, stacking parameters \cite{twisti_angle_dependece_interlayer_dynamics}, and out-of-plane electrical fields \cite{Rivera_field_tuning_interlayerEX,jin2018ultrafast}. 

\begin{figure*}
    \includegraphics[width=1\linewidth]{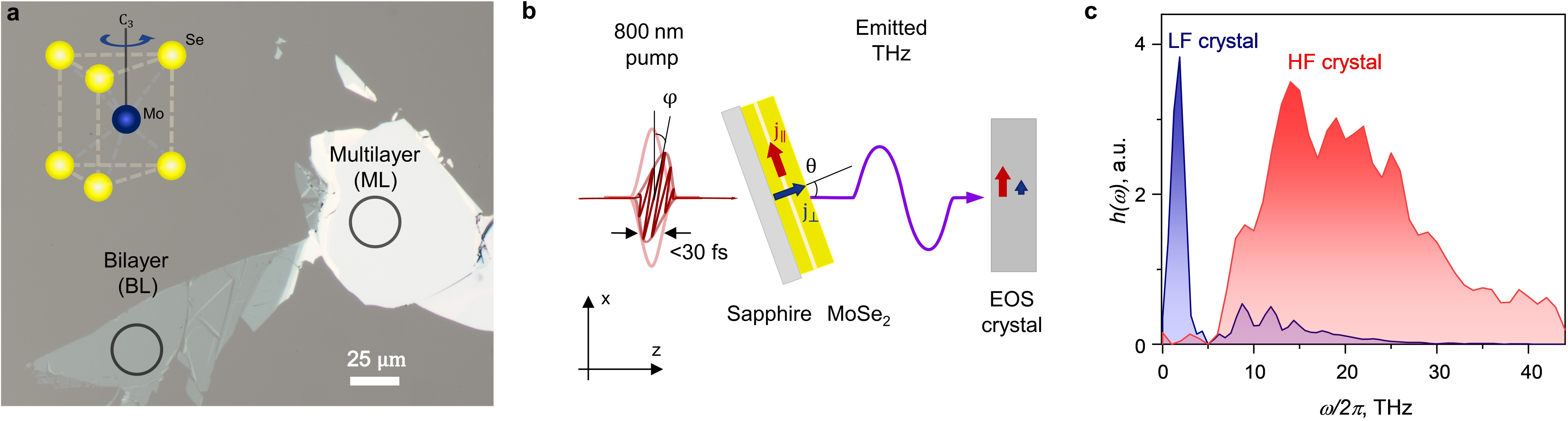}
    \caption{\label{fig:wide}\textbf{Experimental setup} \textbf{a)} Studied bilayer (BL) and multilayer (ML) MoSe$_2$ samples. The laser spot is indicated by black circles. Inset: unit cell of monolayer MoSe$_2$.  \textbf{b)} Measurement geometry. Ultrafast currents in the sample are excited by an optical pump pulse whose polarization is controlled by a half-waveplate (rotation angle $\varphi$). The p-polarized component of the emitted THz wave is detected by electro-optic sampling (EOS). In-plane (red) and out-of-plane (blue) current components are distinguished by tilting ($\theta$) the sample with respect to the optical axis. \textbf{c)} Transfer function $h(\omega)$ of the THz detection that captures THz propagation and EOS with low- and high-frequency (LF and HF) detection crystals ($1\,$mm and $10\,\mu$m thick (110)-cut ZnTe, respectively).}
    \label{fig:1}
\end{figure*}

Most of the studies mentioned above use time-resolved absorption/reflection spectroscopies to probe the dynamics of photoexcited carriers. While a wealth of results has been obtained, these techniques have limitations. For example, it is challenging to resolve the out-of-plane dynamics of a photocurrent. Recently, terahertz (THz) emission spectroscopy has emerged as a useful tool to probe ultrafast current dynamics in thin structures \cite{Ma_Guzelturk_Li_Cao_Shen_Lindenberg_Heinz_2019,Huang_Yao_He_Zhu_Zhang_Bai_Xu_2019,Stein_Fuchs_Stolz_Mittleman_Koch_2020}. In this approach, an ultrafast optical pump pulse is used to excite a photocurrent in the sample. The photocurrent causes the emission of electromagnetic radiation. 
  Duration, direction, and amplitude of this photocurrent define the waveform and amplitude of the detected radiation which is detected by electro-optic sampling (EOS). The advantages of the approach include the high time resolution  \cite{TobiasKa_resolution_Thz_spectr}, excellent sensitivity \cite{Record_sesnsetivity_EOS}, and the ability to detect both in- and out-of-plane currents \cite{Braun_Mussler_Hruban_Konczykowski_Schumann_Wolf_Mnzenberg_Perfetti_Kampfrath_2016}. 

Here, we use THz emission spectroscopy with a very fine time resolution of $\sim30\,$fs to study photocurrents in the prototypical TMDC MoSe$_2$. This time resolution allows us to observe phenomena on time scales that have not been accessible before in TMDCs. We disentangle the dynamics of in-plane and out-of-plane photocurrents by comparing bilayer (BL) and multilayer (ML) MoSe$_2$ samples and by varying the detection angle. Following photoexcitation, we find relatively slow (time scale $\sim0.6\,$ps) out-of-plane photocurrents that are dominant in the ML samples and fast ($\sim50\,$fs) in-plane photocurrents that are dominant in the BL devices. We ascribe the out-of-plane photocurrent in the ML to drift of photoexcited carriers in the surface electric field \cite{Zhang_Hu_Darrow_Auston_1990,huang2017terahertz,Zhang_Huang_Zhao_Zhu_Yao_Zhou_Du_Xu_2017,Si_Huang_Zhao_Zhu_Zhang_Yao_Xu_2018} and the in-plane photocurrent in the BL to a resonant shift current \cite{Zhang_Jin_Yang_Schowalter_1992,Cote_Laman_vanDriel_2002,Sipe_Shkrebtii_2000,Nastos_Sipe_2006,nastos2010optical}. Finally, the large bandwidth of our experiment allows us to  detect oscillations in the THz emission which we attribute to quantum beats between inter- and intra-layer excitonic states in MoSe$_2$  \cite{Gobel_Leo_Damen_Shah_Schmitt-Rink_Schafer_Muller_Kohler_1990,coherent_osc_in_gaas,Brener_Planken_Nuss_Luo_Chuang_Pfeiffer_Leaird_Weiner_1994}.

\section{Setup}

As TMDC sample material, we choose molybdenum diselenide (MoSe$_2$). This material has been shown to have a small density of defects and sharp excitonic peaks \cite{Cadiz_Courtade_Robert_Wang_Shen_Cai_Taniguchi_Watanabe_Carrere_Lagar}. Moreover, its absorption peak is located at $\sim800\,$nm  \cite{About_PLandRaman_of_all_TMDCs}, close to the emission peak of the used Ti:Sapphire ultrafast laser system. At the same time, the material is representative of the entire TMDC family. To study in- and out-of-plane processes in TMDCs, we exfoliated large-area ($>40\times40\,\mu$m) bilayer (BL) and $52\,$nm-thick (Fig. S1) multilayer (ML) MoSe$_2$ samples from bulk crystal (2H-phase)  onto a transparent c-cut sapphire substrate (Fig. 1a). Samples were annealed in vacuum at 200 $^{\circ}$C and characterized using atomic force microscopy (Fig. S2) and photoluminescence spectroscopy (Fig. S3).  

In our measurement setup, the laser-induced photocurrent inside the sample is directly probed by recording the THz radiation generated by this photocurrent using electro-optical sampling (EOS) (Fig. 1b) \cite{phdthesisTHz,Kampfrath_EOS}. The ultrafast laser provides pulses with a width of $30\,$fs, centered at $780\,$nm, with a repetition rate of $80\,$MHz and a pulse energy of $1\,$nJ at the sample position. The beam is split into pump and sampling beams. A half-waveplate is used to control the polarization of the pump beam which is then focused onto the sample by a parabolic mirror resulting in a spot of $\sim25\,\mu$m diameter (black circle in Fig. 1a). The delay between the pump and the sampling beams is set by a precision delay line \cite{Shimada_Kamaraju_Frischkorn_Wolf_Kampfrath_2014}. The emitted THz radiation arising from pump excitation is focused onto a (110)-cut $1\,$mm or $10\,\mu$m thick ZnTe crystal and overlapped with the focused sampling beam. In the EOS process, the initially linearly polarized sampling pulse acquires an ellipticity that is detected using a quarter-waveplate, a polarizing beamsplitter and balanced photodiodes, resulting in the EOS signal $S(t)$ that scales linearly with the p-polarized component of the THz electric field waveform.
Since only the current component perpendicular to the propagation vector of the pump beam ($z$-axis Fig. 1b) emits the detected THz field, we can distinguish between the current projections in and out of the sample plane ($j_\parallel$ and $j_\perp$) by varying the sample tilt angle $\theta$ (Fig. 1b).

In our measurement scheme, the spectral sensitivity of the EOS detection strongly depends on the thickness of the ZnTe crystal \cite{Zhao_Schwagmann_Ospald_Driscoll_Lu_Gossard_Smet_2010}. More precisely, in the frequency domain, the detected EOS signal $S(\omega) = h(\omega)E(\omega)$ is related to the THz electric field $E(\omega)$ just behind the sample through a transfer function $h(\omega)$ which describes the spectral sensitivity of the whole detection process, including the propagation of THz radiation and the detector response (supplementary note 1). In Fig. 1c, we show the measured $h(\omega)$ for $1\,$mm thick (blue curve, termed low frequency (LF)) and $10\,\mu$m thick (red curve, termed high frequency (HF)) ZnTe detection crystals. The LF and HF crystals cover the spectral ranges $0.8$--$2.5\,$THz and $10$--$30\,$THz, respectively. %

Experimental data for LF and HF detection are shown in Fig. 2 and 3, respectively. We first consider the LF data that are dominant in thick multilayer TMDCs before turning our attention to HF processes governing the emission of few-layer samples.

\section{Low-frequency THz emission}

\begin{figure*}
  \begin{minipage}[c]{0.79\textwidth}
    \includegraphics[width=\textwidth]{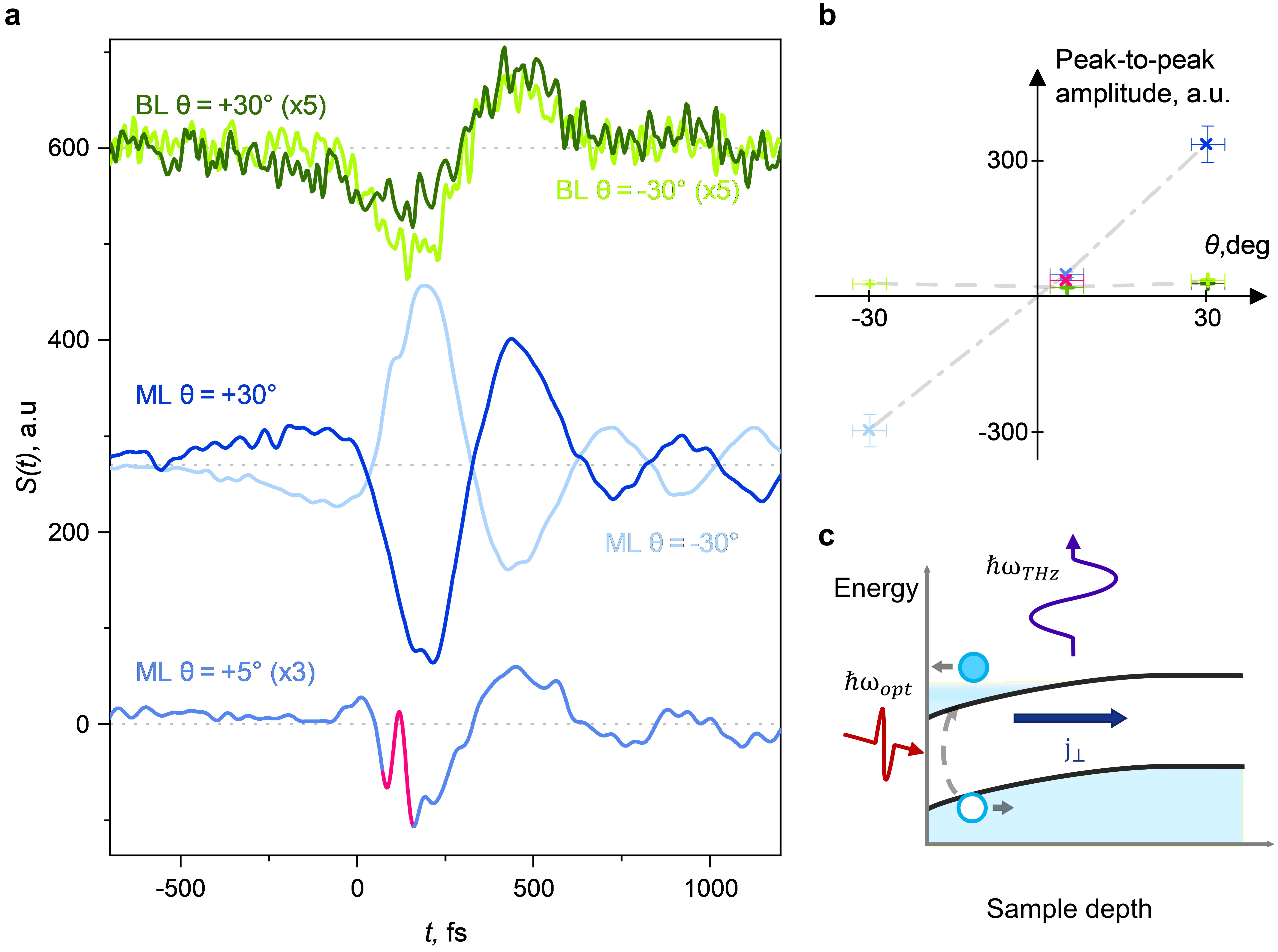}
  \end{minipage}\hfill
  \begin{minipage}[c]{0.21\textwidth}
    \caption{\textbf{Out-of-plane and in-plane LF currents in MoSe$_2$.} \textbf{a)} EOS signal $S(t)$ of the THz electric field emitted by MoSe$_2$ and measured with the LF detector. Data from the bilayer (BL) and the multilayer (ML) samples are shown in green and blue respectively; different colour shades correspond to different sample tilt angles $\theta$. \textbf{b)} Peak-to-peak amplitude of the signal for each sample of a) plotted as a function of the tilt angle $\theta$. The dash and the dash-dotted lines connecting symbols are guides to the eye.  %
     \textbf{c)} Proposed mechanism for out-of-plane THz emission: Photoexcited carriers drift in the opposite direction in the electric field of the surface depletion region, producing an out-of-plane current leading to THz radiation.} \label{fig:03-03}
  \end{minipage}
\end{figure*}

Figure 2a displays a typical EOS signal $S(t)$ of THz pulses emitted by the bilayer region (green) and the multilayer region of the sample (blue) at tilt angles $\theta = +30^{\circ}, -30^{\circ} \text{ and } +5^{\circ} $ measured with the LF detection crystal. 
The EOS signal produced by all samples lasts about $500\,$fs. The residual features seen in Fig. 2a after $500\,$fs are related to THz excitation and reemission of residual water vapor \cite{Beck_Schfer_Klatt_Demsar_Winnerl_Helm_Dekorsy_2010}. %
The dependence of the THz signal on the tilt angle $\theta$ is different between the ML and the BL (shades of blue and green curves in Fig. 2a, respectively, and Fig. S4). The amplitude and the shape of the emission from the BL sample are similar for tilt angles $\theta = +30^{\circ}, -30^{\circ}$, and $+5^{\circ}$ (for angles close to $0^{\circ}$, the back-reflected pump affect the laser mode-locking). In contrast, the emission from the ML flips its sign, while maintaining its amplitude, between $\theta = +30^{\circ}$ and $ -30^{\circ}$. For $\theta = 5^{\circ}$, the ML signal is strongly reduced (note the scale factor of 3 in Fig. 2a). In addition, a high-frequency feature becomes apparent in the signal at $t \approx 0$ (pink trace in Fig. 2a). The broad spectrum ($5\,$--$20\,$THz) of this feature (Fig. S5) coincides with spectral regions where the sensitivity of the LF crystal is low (blue curve in Fig. 1c).

The tilt angle dependence of the peak-to-peak amplitudes of the emitted THz pulse for both ML and BL samples is summarized in Fig. 2b. The dependence suggests that the signal in the BL is primarily produced by in-plane photocurrents, whereas the signal in the ML samples predominantly stems from out-of-plane photocurrents. Indeed, in-plane currents $j_{\parallel}$ should produce a signal proportional to $ \cos\theta$ without a change in sign. This matches the observed behavior for the BL sample (green symbols in Fig. 2b). In contrast, THz emission from out-plane currents should be proportional to $\sin\theta$ and therefore vanish at $\theta \approx 0$. This behavior is observed for the ML THz emission (blue symbols in Fig. 2b), except for the sharp feature in Fig. 2 (pink symbols in Fig. 2b, see high-pass filtered data in Fig. S5). Finally, the fact that the sharp feature in the ML signal (the pink trace in Fig. 2a) persists at $\theta = 5^{\circ}$ suggests that this feature has an in-plane character. This assignment is also supported by a symmetry analysis (Fig. S6).

What are the origins of the observed in-plane and out-of-plane currents in the multilayer MoSe$_2$? Similar out-of-plane emission has been previously observed in other thick TMDCs and ascribed to a drift of photoexcited charge carriers in the built-in electric field appearing close to the surface of the sample due to the surface charged states (Fig. 2c) \cite{huang2017terahertz,Zhang_Huang_Zhao_Zhu_Yao_Zhou_Du_Xu_2017,Si_Huang_Zhao_Zhu_Zhang_Yao_Xu_2018}. Indeed, such a mechanism is out-of-plane, is very weak or absent in the BL sample, and should cause THz emission on a picosecond timescale \cite{Johnston_Whittaker_Corchia_Davies_Linfield_2002,Heyman_Coates_Reinhardt_Strasser_2003}. We note that the emission from another representative TMDC, multilayer WSe$_2$ has similar amplitude and dynamics (Fig. S7). This is consistent with measurements of that material by others \cite{Si_Huang_Zhao_Zhu_Zhang_Yao_Xu_2018,Huang_Yao_He_Zhu_Zhang_Bai_Xu_2019} and suggests that the proposed emission mechanism is generic for the TMDC family. %
At the first glance, it would appear that the in-plane emission occurs at two different timescales~---~around $500\,$fs for the BL (green curves in Fig. 2a) and $50 \,$fs for the ML (the pink trace in Fig. 2a). However, the spectral analysis of the latter fast feature in the ML emission reveals significant contribution from outside of the high-sensitivity frequency region (up to $2.5\,$THz) of the LF crystal (Fig. S5). Because of that, we hypothesize that the apparently "slow" in-plane process in the BL may also originate from an intrinsically "fast" process. We therefore switch to a HF detection crystal (red in Fig. 1c) to further test this hypothesis.

\section{High-frequency THz emission}

\begin{figure*}
    \includegraphics[width=0.97\linewidth]{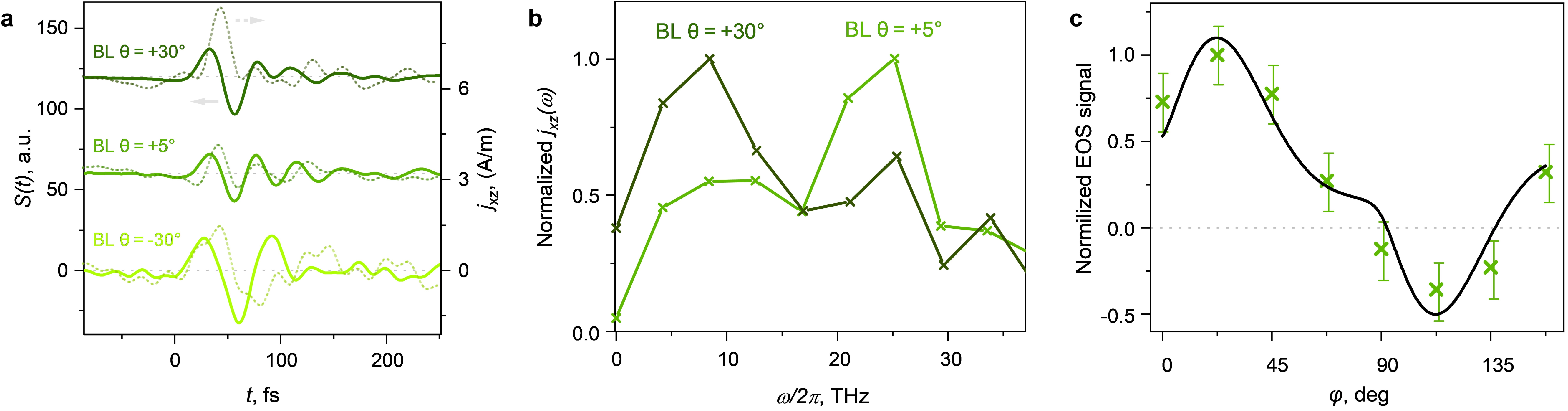}
    \caption{\label{fig:wide}\textbf{In-plane HF currents in MoSe$_2$ a)} EOS signal of the THz electric field measured with the HF detector (solid lines) and extracted charge currents $j$ (dotted lines) in the bilayer MoSe$_2$ sample for tilt angles $\theta=+30^{\circ}, +5^{\circ}, -30^{\circ}$. The relatively weak dependence of amplitude and temporal profile on $\theta$ suggests the in-plane character of the ultrafast current. \ \textbf{b)} Fourier amplitude spectrum of the extracted charge current. A peak at around $8\,$THz is ascribed to a resonant shift current in the sample. A peak at $\sim23\,$THz corresponds to the decaying oscillations seen in a) for $\theta=5^{\circ}$. The frequency of these oscillations suggests that they may arise from quantum beating of inter- and intra-layer excitons in MoSe$_2$. \textbf{c)} Dependence of the normalized amplitude of the EOS signal $S(t)$ on the tilt angle of the pump polarization plane ($\varphi$) for $\theta=5^{\circ}$ (points). The polarization dependence is consistent with the assumption that the THz signal is due to second-order optical effects (solid line).}
    \label{fig:3}
\end{figure*}

The emission from MoSe$_2$ measured at tilt angles $\theta=+30^{\circ},+5^{\circ},-30^{\circ}$ using the $10\,\mu$m thick ZnTe crystal sensitive to high-frequencies is shown in Fig. 3a as solid lines for the BL sample and in the Fig. S8 for the ML sample. The dashed lines in Fig. 3a represent the electrical current $j_{xz}(t)$ inside of the sample extracted from EOS signal $S(t)$ following supplementary note 2. The BL emission features a bipolar swing followed by decaying oscillations that are analyzed later. The timescale of the swing is approximately $50\,$fs; its amplitude is similar for all tilt angles $\theta=+30^{\circ},+5^{\circ},-30^{\circ}$ (Fig. 3a). This suggests the in-plane current nature of the high-frequency emission in the sample. In the ML samples, the tilt angle dependence is more complex (Fig. S8), however the EOS signal does not flip sign, which is expected for in-plane generating current. Therefore, the HF emission in all samples has an in-plane origin.

Several distinct physical mechanisms contribute to HF in-plane emission in bilayer MoSe$_2$. From a phenomenological viewpoint, we first note that the currents observed here are non-linear second-order optical effects (SOEs). SOEs are expected to produce emission at a difference frequency $\omega=\omega_1-\omega_2$, for any two frequencies $\omega_1$ and $\omega_2$ inside the pump pulse spectrum \cite{Nastos_Sipe_2006,Fregoso_2019}. In the case of our $30\,$fs-long excitations, we expect the emission with a bandwidth of $\sim33\,$THz, the inverse of the pump pulse duration. This is consistent with the experimentally observed bandwidth from $\sim0.3\,$THz up to $30\,$THz in the BL MoSe$_2$ (see Fig. 3b and Fig. S4 for Fourier transform of green curves in Fig. 2a). The lower amplitude of the HF components in spectrum of the BL emission at $\theta = -30^{\circ}$ is related to variation in sensitivity of the detection at high frequency. In general, SOEs appear in media which lack inversion symmetry \cite{Nastos_Sipe_2006,Sipe_Shkrebtii_2000}. While even-layered MoSe$_2$ is inversion-symmetric, inversion symmetry is broken in each individual layer of the structure \cite{Manzeli_Ovchinnikov_Pasquier_Yazyev_Kis_2017} (Fig. 1a, Inset). The bending of the MoSe$_2$ band structure at interfaces, responsible for ML emission (Fig. 2c), causes an asymmetry between the top and the bottom layers of the stacks \cite{Si_Huang_Zhao_Zhu_Zhang_Yao_Xu_2018,Zhang_Huang_Zhao_Zhu_Yao_Zhou_Du_Xu_2017,Steinberg_Gardner_Lee_Jarillo-Herrero_2010,McIver_Hsieh_Drapcho_Torchinsky_Gardner_Lee_Gedik_2012}. %
We obtain further evidence for the contribution of SOEs by analyzing the dependence of the emitted THz from the BL on the polarization angle of the pump pulse $\varphi$ (Fig. 3c). We observe $\sin(2\varphi)$-like behavior in the emitted signal (points in Fig. 3c), where $\varphi$ is the tilt of the polarization plane of the initially p-polarized pump beam. This dependence is expected for SOEs as their efficiency depends on the respective alignment between the electrical field of the pump and crystallographic axes of the sample  \cite{Nastos_Sipe_2006,Braun_Mussler_Hruban_Konczykowski_Schumann_Wolf_Mnzenberg_Perfetti_Kampfrath_2016,Huang_Zhu_Zhao_Guo_Ren_Bai_Xu_2017}. In the supplementary note 2, we derive an analytical expression for the angle-dependence of the emitted field in the BL MoSe$_2$. A good match of this expression (black line in Fig. 3c) to the experimental data (crosses in Fig. 3c) confirms second-order optical effects as dominant processes.

 
To discuss the microscopic origins of the photocurrents observed here, we note that, in general, multiple distinct SOE effects may contribute in our samples, the most notable of which are injection currents and shift currents \cite{Zhang_Jin_Yang_Schowalter_1992,Cote_Laman_vanDriel_2002,Sipe_Shkrebtii_2000,Nastos_Sipe_2006,nastos2010optical}. %
  The injection currents are only expected under illumination with circularly polarized light, while we use only linearly polarized pump pulses \cite{Nastos_Sipe_2006,Fregoso_2019}. %
The off-resonant shift current (also referred to as "optical rectification") is expected to generate order of magnitude smaller currents \cite{Braun_Mussler_Hruban_Konczykowski_Schumann_Wolf_Mnzenberg_Perfetti_Kampfrath_2016, Ferguson_Zhang_2002} than the remaining SOE mechanism: the resonant shift current. The latter current arises from a spatial shift of the center of mass of the electron density within a unit cell when excited from the valence to the conduction bands \cite{Braun_Mussler_Hruban_Konczykowski_Schumann_Wolf_Mnzenberg_Perfetti_Kampfrath_2016,Fregoso_2019}. Indeed, the photon energy of the pump pulse ($1.51\,$--$1.70\,$eV) is in resonance with excitonic transitions in the BL MoSe$_2$ ($1.53\,$eV, Fig. S3). Moreover, a swing from positive to negative photocurrent, most prominently seen in Fig. 3a at $\theta = 30^{\circ}$, is a temporal feature expected from a shift current but not from other SOE mechanisms \cite{Braun_Mussler_Hruban_Konczykowski_Schumann_Wolf_Mnzenberg_Perfetti_Kampfrath_2016}. Therefore, we assign the THz emission from the BL MoSe$_2$ to the resonant shift current.

We finally address the oscillations in the EOS signal $S(t)$ that are especially prominent in the BL signal at $\theta = 5^{\circ}$ tilt (Fig. 3a after $\sim75\,$fs). The same feature is evident in the extracted current for bilayer MoSe$_2$ (dashed lines in Fig. 3a). %

The spectral content of the oscillations is distinct from that of other types of emission we observed in the ML or the BL samples. The Fourier transform with a narrow square window of $240\,$fs is shown in Fig. 3b (see the analysis in supplementary Fig. S9). It features two distinct peaks: a low-frequency (centered around $8\,$THz) peak, assigned to the resonant shift current considered earlier, and a high-frequency peak (around $23\,$THz). Note that the transfer function of the used detection scheme (red in Fig. 1c) is flat around the center of this peak. Therefore, this peak is not due to the limited spectral response of the detector. 
 Relative intensity of the high-frequency peak depends on the tilt angle $\theta$ and matches visibility of the oscillations. 

We suggest that the high frequency ($23\pm3\,$THz) and fast decay constant of $55\,$fs (Fig. S10) of the photocurrent oscillations are related to quantum beats between two states separated by the corresponding energy ($95\pm15$meV) and coherently excited by the pump. This type of emission was first observed with heavy and light hole states in GaAs quantum wells at low temperature \cite{coherent_osc_in_gaas,Brener_Planken_Nuss_Luo_Chuang_Pfeiffer_Leaird_Weiner_1994} and theoretically predicted to be efficient for in-plane and out-of-plane excitons  \cite{kyriienko2013superradiant,kristinsson2013continuous,kristinsson2014terahertz}, which are found in TMDCs \cite{Rivera_field_tuning_interlayerEX,Inter_Ex_MoS2WS2}. Moreover,  quantum beats between anisotropic excitons were recently observed in time-resolved differential transmittance of atomically thin ReS$_2$ TMDCs at low temperature \cite{Sim_Lee_Trifonov_Kim_Cha_Sung_Cho_Shim_Jo_Choi_2018}. A pair of excitonic states in the bilayer MoSe$_2$ corresponding to observed quantum beats should i) fall within the pump spectral bandwidth $1.51\,$--$1.70\,$eV, ii) have the energy separation $\sim95\pm15\,$meV, iii) different dipole moments, and iv) comparable oscillator strengths. The following excitonic states lie within the bandwidth of the pump: neutral ground state intralayer exciton (A), excited states of this exciton (An), charged exciton (trion T), biexciton (X), and interlayer exciton (I) \cite{Biexcitons,interlayer_excitons_in_MoSe2_BL}. Of these, the only states matching the required energy separation is the A-I pair with the energy difference $\Delta E_{A1-I}\approx82\,$meV. We note that this energy difference is extrapolated from the low-temperature measurements \cite{interlayer_excitons_in_MoSe2_BL,Gerber_Courtade_Shree_Robert_Taniguchi_Watanabe_Balocchi_Renucci_Lagarde_Marie_2019}, where these peaks are clearly visible, to room temperature. In addition, the two states A and I have comparable oscillator strengths \cite{interlayer_excitons_in_MoSe2_BL,Gerber_Courtade_Shree_Robert_Taniguchi_Watanabe_Balocchi_Renucci_Lagarde_Marie_2019} and different dipole moments, in-plane for A \cite{schuller2013orientation} and out-of-plane for I \cite{wang2017electrical}. Finally, the decay time of $55\,$fs (fig. S10) matches the dephasing time for excitonic species extrapolated from low to room temperature \cite{FWM_Tr_NX,FWM_Tr,fwm_NX}. Therefore, we propose that quantum beats between intra- and inter-layer excitonic species are related the observed oscillations in THz emission.

To summarize, we observed THz photocurrents in bilayer (BL) and multilayer (ML) MoSe$_2$ in both low-frequency ($0.8$--$2.5\,$THz) and high-frequency ($10$--$30\,$THz) spectral regions using ultrafast terahertz emission spectroscopy. The low-frequency emission from the multilayer sample features a radiation pattern typical for an out-of-plane charge flow. We attribute this emission to drift of photogenerated carriers in the surface electric filed. In contrast, the angular distribution of the THz emission from the BL sample is consistent with in-plane photocurrents. By analyzing the polarization dependence and amplitude of this emission at high frequency, we are led to ascribe it to a resonant shift current. Furthermore, THz emission from the BL sample features oscillations of the emitted field. Frequency and decay time of these oscillations are consistent with quantum beats between coherently excited inter- and intra-layer excitons as generation process. Our findings provide a picture of photocurrent processes following ultrafast optical excitation of TMDCs. Finally, quantum beats may enable novel wavelength-tunable room temperature THz emitters once electrical tunability is demonstrated and THz emission intensity is improved.

We thank S. Heeg for great comments on the paper, A. Kumar for useful discussions, and M. Borchert for help with the experiment. We acknowledge the German Research Foundation (DFG) for financial support through the Collaborative Research Center TRR 227 Ultrafast Spin Dynamics (projects B08, A05 and B02) and the European Research Council for funding through the H2020-CoG project TERAMAG (grant no. 681917).

\bibliography{apssamp}

\providecommand{\noopsort}[1]{}\providecommand{\singleletter}[1]{#1}%
\begin{thebibliography}{63}%
\makeatletter
\providecommand \@ifxundefined [1]{%
 \@ifx{#1\undefined}
}%
\providecommand \@ifnum [1]{%
 \ifnum #1\expandafter \@firstoftwo
 \else \expandafter \@secondoftwo
 \fi
}%
\providecommand \@ifx [1]{%
 \ifx #1\expandafter \@firstoftwo
 \else \expandafter \@secondoftwo
 \fi
}%
\providecommand \natexlab [1]{#1}%
\providecommand \enquote  [1]{``#1''}%
\providecommand \bibnamefont  [1]{#1}%
\providecommand \bibfnamefont [1]{#1}%
\providecommand \citenamefont [1]{#1}%
\providecommand \href@noop [0]{\@secondoftwo}%
\providecommand \href [0]{\begingroup \@sanitize@url \@href}%
\providecommand \@href[1]{\@@startlink{#1}\@@href}%
\providecommand \@@href[1]{\endgroup#1\@@endlink}%
\providecommand \@sanitize@url [0]{\catcode `\\12\catcode `\$12\catcode
  `\&12\catcode `\#12\catcode `\^12\catcode `\_12\catcode `\%12\relax}%
\providecommand \@@startlink[1]{}%
\providecommand \@@endlink[0]{}%
\providecommand \url  [0]{\begingroup\@sanitize@url \@url }%
\providecommand \@url [1]{\endgroup\@href {#1}{\urlprefix }}%
\providecommand \urlprefix  [0]{URL }%
\providecommand \Eprint [0]{\href }%
\providecommand \doibase [0]{https://doi.org/}%
\providecommand \selectlanguage [0]{\@gobble}%
\providecommand \bibinfo  [0]{\@secondoftwo}%
\providecommand \bibfield  [0]{\@secondoftwo}%
\providecommand \translation [1]{[#1]}%
\providecommand \BibitemOpen [0]{}%
\providecommand \bibitemStop [0]{}%
\providecommand \bibitemNoStop [0]{.\EOS\space}%
\providecommand \EOS [0]{\spacefactor3000\relax}%
\providecommand \BibitemShut  [1]{\csname bibitem#1\endcsname}%
\let\auto@bib@innerbib\@empty
\bibitem [{\citenamefont {Amani}\ \emph {et~al.}(2015)\citenamefont {Amani},
  \citenamefont {Lien}, \citenamefont {Kiriya}, \citenamefont {Xiao},
  \citenamefont {Azcatl}, \citenamefont {Noh}, \citenamefont {Madhvapathy},
  \citenamefont {Addou}, \citenamefont {KC}, \citenamefont {Dubey},\ and\
  \citenamefont {et~al.}}]{near_unity_yeald_MoS2}%
  \BibitemOpen
  \bibfield  {author} {\bibinfo {author} {\bibfnamefont {M.}~\bibnamefont
  {Amani}}, \bibinfo {author} {\bibfnamefont {D.-H.}\ \bibnamefont {Lien}},
  \bibinfo {author} {\bibfnamefont {D.}~\bibnamefont {Kiriya}}, \bibinfo
  {author} {\bibfnamefont {J.}~\bibnamefont {Xiao}}, \bibinfo {author}
  {\bibfnamefont {A.}~\bibnamefont {Azcatl}}, \bibinfo {author} {\bibfnamefont
  {J.}~\bibnamefont {Noh}}, \bibinfo {author} {\bibfnamefont {S.~R.}\
  \bibnamefont {Madhvapathy}}, \bibinfo {author} {\bibfnamefont
  {R.}~\bibnamefont {Addou}}, \bibinfo {author} {\bibfnamefont
  {S.}~\bibnamefont {KC}}, \bibinfo {author} {\bibfnamefont {M.}~\bibnamefont
  {Dubey}},\ and\ \bibinfo {author} {\bibnamefont {et~al.}},\ }\bibfield
  {title} {\bibinfo {title} {Near-unity photoluminescence quantum yield in
  mos2},\ }\href {https://doi.org/10.1126/science.aad2114} {\bibfield
  {journal} {\bibinfo  {journal} {Science}\ }\textbf {\bibinfo {volume}
  {350}},\ \bibinfo {pages} {1065–1068} (\bibinfo {year} {2015})}\BibitemShut
  {NoStop}%
\bibitem [{\citenamefont {Massicotte}\ \emph {et~al.}(2018)\citenamefont
  {Massicotte}, \citenamefont {Vialla}, \citenamefont {Schmidt}, \citenamefont
  {Lundeberg}, \citenamefont {Latini}, \citenamefont {Haastrup}, \citenamefont
  {Danovich}, \citenamefont {Davydovskaya}, \citenamefont {Watanabe},
  \citenamefont {Taniguchi},\ and\ \citenamefont {et~al.}}]{dynamics_Ex_WSe2}%
  \BibitemOpen
  \bibfield  {author} {\bibinfo {author} {\bibfnamefont {M.}~\bibnamefont
  {Massicotte}}, \bibinfo {author} {\bibfnamefont {F.}~\bibnamefont {Vialla}},
  \bibinfo {author} {\bibfnamefont {P.}~\bibnamefont {Schmidt}}, \bibinfo
  {author} {\bibfnamefont {M.~B.}\ \bibnamefont {Lundeberg}}, \bibinfo {author}
  {\bibfnamefont {S.}~\bibnamefont {Latini}}, \bibinfo {author} {\bibfnamefont
  {S.}~\bibnamefont {Haastrup}}, \bibinfo {author} {\bibfnamefont
  {M.}~\bibnamefont {Danovich}}, \bibinfo {author} {\bibfnamefont
  {D.}~\bibnamefont {Davydovskaya}}, \bibinfo {author} {\bibfnamefont
  {K.}~\bibnamefont {Watanabe}}, \bibinfo {author} {\bibfnamefont
  {T.}~\bibnamefont {Taniguchi}},\ and\ \bibinfo {author} {\bibnamefont
  {et~al.}},\ }\bibfield  {title} {\bibinfo {title} {Dissociation of
  two-dimensional excitons in monolayer wse2},\ }\bibfield  {journal} {\bibinfo
   {journal} {Nature Communications}\ }\textbf {\bibinfo {volume} {9}},\ \href
  {https://doi.org/10.1038/s41467-018-03864-y} {10.1038/s41467-018-03864-y}
  (\bibinfo {year} {2018})\BibitemShut {NoStop}%
\bibitem [{\citenamefont {Godde}\ \emph {et~al.}(2016)\citenamefont {Godde},
  \citenamefont {Schmidt}, \citenamefont {Schmutzler}, \citenamefont {Aßmann},
  \citenamefont {Debus}, \citenamefont {Withers}, \citenamefont {Alexeev},
  \citenamefont {Del Pozo-Zamudio}, \citenamefont {Skrypka}, \citenamefont
  {Novoselov},\ and\ \citenamefont {et~al.}}]{Tr_Ex_MoSe2_and_WSe2}%
  \BibitemOpen
  \bibfield  {author} {\bibinfo {author} {\bibfnamefont {T.}~\bibnamefont
  {Godde}}, \bibinfo {author} {\bibfnamefont {D.}~\bibnamefont {Schmidt}},
  \bibinfo {author} {\bibfnamefont {J.}~\bibnamefont {Schmutzler}}, \bibinfo
  {author} {\bibfnamefont {M.}~\bibnamefont {Aßmann}}, \bibinfo {author}
  {\bibfnamefont {J.}~\bibnamefont {Debus}}, \bibinfo {author} {\bibfnamefont
  {F.}~\bibnamefont {Withers}}, \bibinfo {author} {\bibfnamefont {E.~M.}\
  \bibnamefont {Alexeev}}, \bibinfo {author} {\bibfnamefont {O.}~\bibnamefont
  {Del Pozo-Zamudio}}, \bibinfo {author} {\bibfnamefont {O.~V.}\ \bibnamefont
  {Skrypka}}, \bibinfo {author} {\bibfnamefont {K.~S.}\ \bibnamefont
  {Novoselov}},\ and\ \bibinfo {author} {\bibnamefont {et~al.}},\ }\bibfield
  {title} {\bibinfo {title} {Exciton and trion dynamics in atomically thin
  mose2 and wse2: Effect of localization},\ }\bibfield  {journal} {\bibinfo
  {journal} {Physical Review B}\ }\textbf {\bibinfo {volume} {94}},\ \href
  {https://doi.org/10.1103/physrevb.94.165301} {10.1103/physrevb.94.165301}
  (\bibinfo {year} {2016})\BibitemShut {NoStop}%
\bibitem [{\citenamefont {Li}\ \emph {et~al.}(2014)\citenamefont {Li},
  \citenamefont {Chernikov}, \citenamefont {Zhang}, \citenamefont {Rigosi},
  \citenamefont {Hill}, \citenamefont {van~der Zande}, \citenamefont {Chenet},
  \citenamefont {Shih}, \citenamefont {Hone},\ and\ \citenamefont
  {Heinz}}]{absorption_spectrum_TMDCs}%
  \BibitemOpen
  \bibfield  {author} {\bibinfo {author} {\bibfnamefont {Y.}~\bibnamefont
  {Li}}, \bibinfo {author} {\bibfnamefont {A.}~\bibnamefont {Chernikov}},
  \bibinfo {author} {\bibfnamefont {X.}~\bibnamefont {Zhang}}, \bibinfo
  {author} {\bibfnamefont {A.}~\bibnamefont {Rigosi}}, \bibinfo {author}
  {\bibfnamefont {H.~M.}\ \bibnamefont {Hill}}, \bibinfo {author}
  {\bibfnamefont {A.~M.}\ \bibnamefont {van~der Zande}}, \bibinfo {author}
  {\bibfnamefont {D.~A.}\ \bibnamefont {Chenet}}, \bibinfo {author}
  {\bibfnamefont {E.-M.}\ \bibnamefont {Shih}}, \bibinfo {author}
  {\bibfnamefont {J.}~\bibnamefont {Hone}},\ and\ \bibinfo {author}
  {\bibfnamefont {T.~F.}\ \bibnamefont {Heinz}},\ }\bibfield  {title} {\bibinfo
  {title} {Measurement of the optical dielectric function of monolayer
  transition-metal dichalcogenides: Mos2, mose2, ws2, and wse2},\ }\bibfield
  {journal} {\bibinfo  {journal} {Physical Review B}\ }\textbf {\bibinfo
  {volume} {90}},\ \href {https://doi.org/10.1103/physrevb.90.205422}
  {10.1103/physrevb.90.205422} (\bibinfo {year} {2014})\BibitemShut {NoStop}%
\bibitem [{\citenamefont {Klots}\ \emph {et~al.}(2014)\citenamefont {Klots},
  \citenamefont {Newaz}, \citenamefont {Wang}, \citenamefont {Prasai},
  \citenamefont {Krzyzanowska}, \citenamefont {Lin}, \citenamefont {Caudel},
  \citenamefont {Ghimire}, \citenamefont {Yan}, \citenamefont {Ivanov},\ and\
  \citenamefont
  {et~al.}}]{Klots_Newaz_Wang_Prasai_Krzyzanowska_Lin_Caudel_Ghimire_Yan_Ivanov}%
  \BibitemOpen
  \bibfield  {author} {\bibinfo {author} {\bibfnamefont {A.~R.}\ \bibnamefont
  {Klots}}, \bibinfo {author} {\bibfnamefont {A.~K.~M.}\ \bibnamefont {Newaz}},
  \bibinfo {author} {\bibfnamefont {B.}~\bibnamefont {Wang}}, \bibinfo {author}
  {\bibfnamefont {D.}~\bibnamefont {Prasai}}, \bibinfo {author} {\bibfnamefont
  {H.}~\bibnamefont {Krzyzanowska}}, \bibinfo {author} {\bibfnamefont
  {J.}~\bibnamefont {Lin}}, \bibinfo {author} {\bibfnamefont {D.}~\bibnamefont
  {Caudel}}, \bibinfo {author} {\bibfnamefont {N.~J.}\ \bibnamefont {Ghimire}},
  \bibinfo {author} {\bibfnamefont {J.}~\bibnamefont {Yan}}, \bibinfo {author}
  {\bibfnamefont {B.~L.}\ \bibnamefont {Ivanov}},\ and\ \bibinfo {author}
  {\bibnamefont {et~al.}},\ }\bibfield  {title} {\bibinfo {title} {Probing
  excitonic states in suspended two-dimensional semiconductors by photocurrent
  spectroscopy},\ }\bibfield  {journal} {\bibinfo  {journal} {Scientific
  Reports}\ }\textbf {\bibinfo {volume} {4}},\ \href
  {https://doi.org/10.1038/srep06608} {10.1038/srep06608} (\bibinfo {year}
  {2014})\BibitemShut {NoStop}%
\bibitem [{\citenamefont {Chernikov}\ \emph {et~al.}(2014)\citenamefont
  {Chernikov}, \citenamefont {Berkelbach}, \citenamefont {Hill}, \citenamefont
  {Rigosi}, \citenamefont {Li}, \citenamefont {Aslan}, \citenamefont
  {Reichman}, \citenamefont {Hybertsen},\ and\ \citenamefont
  {Heinz}}]{Chernikov_Berkelbach_Hill_Rigosi_Li_Aslan_Reichman_Hybertsen_Heinz_2014}%
  \BibitemOpen
  \bibfield  {author} {\bibinfo {author} {\bibfnamefont {A.}~\bibnamefont
  {Chernikov}}, \bibinfo {author} {\bibfnamefont {T.~C.}\ \bibnamefont
  {Berkelbach}}, \bibinfo {author} {\bibfnamefont {H.~M.}\ \bibnamefont
  {Hill}}, \bibinfo {author} {\bibfnamefont {A.}~\bibnamefont {Rigosi}},
  \bibinfo {author} {\bibfnamefont {Y.}~\bibnamefont {Li}}, \bibinfo {author}
  {\bibfnamefont {O.~B.}\ \bibnamefont {Aslan}}, \bibinfo {author}
  {\bibfnamefont {D.~R.}\ \bibnamefont {Reichman}}, \bibinfo {author}
  {\bibfnamefont {M.~S.}\ \bibnamefont {Hybertsen}},\ and\ \bibinfo {author}
  {\bibfnamefont {T.~F.}\ \bibnamefont {Heinz}},\ }\bibfield  {title} {\bibinfo
  {title} {Exciton binding energy and nonhydrogenic rydberg series in monolayer
  ws2},\ }\bibfield  {journal} {\bibinfo  {journal} {Physical Review Letters}\
  }\textbf {\bibinfo {volume} {113}},\ \href
  {https://doi.org/10.1103/physrevlett.113.076802}
  {10.1103/physrevlett.113.076802} (\bibinfo {year} {2014})\BibitemShut
  {NoStop}%
\bibitem [{\citenamefont {Klots}\ \emph {et~al.}(2018)\citenamefont {Klots},
  \citenamefont {Weintrub}, \citenamefont {Prasai}, \citenamefont {Kidd},
  \citenamefont {Varga}, \citenamefont {Velizhanin},\ and\ \citenamefont
  {Bolotin}}]{Klots_Weintrub_Prasai_Kidd_Varga_Velizhanin_Bolotin_2018}%
  \BibitemOpen
  \bibfield  {author} {\bibinfo {author} {\bibfnamefont {A.~R.}\ \bibnamefont
  {Klots}}, \bibinfo {author} {\bibfnamefont {B.}~\bibnamefont {Weintrub}},
  \bibinfo {author} {\bibfnamefont {D.}~\bibnamefont {Prasai}}, \bibinfo
  {author} {\bibfnamefont {D.}~\bibnamefont {Kidd}}, \bibinfo {author}
  {\bibfnamefont {K.}~\bibnamefont {Varga}}, \bibinfo {author} {\bibfnamefont
  {K.~A.}\ \bibnamefont {Velizhanin}},\ and\ \bibinfo {author} {\bibfnamefont
  {K.~I.}\ \bibnamefont {Bolotin}},\ }\bibfield  {title} {\bibinfo {title}
  {Controlled dynamic screening of excitonic complexes in 2d semiconductors},\
  }\bibfield  {journal} {\bibinfo  {journal} {Scientific Reports}\ }\textbf
  {\bibinfo {volume} {8}},\ \href {https://doi.org/10.1038/s41598-017-18803-y}
  {10.1038/s41598-017-18803-y} (\bibinfo {year} {2018})\BibitemShut {NoStop}%
\bibitem [{\citenamefont {Greben}\ \emph {et~al.}(2020)\citenamefont {Greben},
  \citenamefont {Arora}, \citenamefont {Harats},\ and\ \citenamefont
  {Bolotin}}]{Greben_Arora_Harats_Bolotin_2020}%
  \BibitemOpen
  \bibfield  {author} {\bibinfo {author} {\bibfnamefont {K.}~\bibnamefont
  {Greben}}, \bibinfo {author} {\bibfnamefont {S.}~\bibnamefont {Arora}},
  \bibinfo {author} {\bibfnamefont {M.~G.}\ \bibnamefont {Harats}},\ and\
  \bibinfo {author} {\bibfnamefont {K.~I.}\ \bibnamefont {Bolotin}},\
  }\bibfield  {title} {\bibinfo {title} {Intrinsic and extrinsic defect-related
  excitons in tmdcs},\ }\href {https://doi.org/10.1021/acs.nanolett.9b05323}
  {\bibfield  {journal} {\bibinfo  {journal} {Nano Letters}\ }\textbf {\bibinfo
  {volume} {20}},\ \bibinfo {pages} {2544–2550} (\bibinfo {year}
  {2020})}\BibitemShut {NoStop}%
\bibitem [{\citenamefont {Poellmann}\ \emph {et~al.}(2015)\citenamefont
  {Poellmann}, \citenamefont {Steinleitner}, \citenamefont {Leierseder},
  \citenamefont {Nagler}, \citenamefont {Plechinger}, \citenamefont {Porer},
  \citenamefont {Bratschitsch}, \citenamefont {Schüller}, \citenamefont
  {Korn},\ and\ \citenamefont
  {Huber}}]{Poellmann_Steinleitner_Leierseder_Nagler_Plechinger_Porer_Bratschitsch_Schller_Korn_Huber_2015}%
  \BibitemOpen
  \bibfield  {author} {\bibinfo {author} {\bibfnamefont {C.}~\bibnamefont
  {Poellmann}}, \bibinfo {author} {\bibfnamefont {P.}~\bibnamefont
  {Steinleitner}}, \bibinfo {author} {\bibfnamefont {U.}~\bibnamefont
  {Leierseder}}, \bibinfo {author} {\bibfnamefont {P.}~\bibnamefont {Nagler}},
  \bibinfo {author} {\bibfnamefont {G.}~\bibnamefont {Plechinger}}, \bibinfo
  {author} {\bibfnamefont {M.}~\bibnamefont {Porer}}, \bibinfo {author}
  {\bibfnamefont {R.}~\bibnamefont {Bratschitsch}}, \bibinfo {author}
  {\bibfnamefont {C.}~\bibnamefont {Schüller}}, \bibinfo {author}
  {\bibfnamefont {T.}~\bibnamefont {Korn}},\ and\ \bibinfo {author}
  {\bibfnamefont {R.}~\bibnamefont {Huber}},\ }\bibfield  {title} {\bibinfo
  {title} {Resonant internal quantum transitions and femtosecond radiative
  decay of excitons in monolayer wse2},\ }\href
  {https://doi.org/10.1038/nmat4356} {\bibfield  {journal} {\bibinfo  {journal}
  {Nature Materials}\ }\textbf {\bibinfo {volume} {14}},\ \bibinfo {pages}
  {889–893} (\bibinfo {year} {2015})}\BibitemShut {NoStop}%
\bibitem [{\citenamefont {Singh}\ \emph {et~al.}(2016)\citenamefont {Singh},
  \citenamefont {Moody}, \citenamefont {Tran}, \citenamefont {Scott},
  \citenamefont {Overbeck}, \citenamefont {Berghäuser}, \citenamefont
  {Schaibley}, \citenamefont {Seifert}, \citenamefont {Pleskot}, \citenamefont
  {Gabor},\ and\ \citenamefont {et~al.}}]{Singh_Moody_Tran_Scott_Overbeck_B}%
  \BibitemOpen
  \bibfield  {author} {\bibinfo {author} {\bibfnamefont {A.}~\bibnamefont
  {Singh}}, \bibinfo {author} {\bibfnamefont {G.}~\bibnamefont {Moody}},
  \bibinfo {author} {\bibfnamefont {K.}~\bibnamefont {Tran}}, \bibinfo {author}
  {\bibfnamefont {M.~E.}\ \bibnamefont {Scott}}, \bibinfo {author}
  {\bibfnamefont {V.}~\bibnamefont {Overbeck}}, \bibinfo {author}
  {\bibfnamefont {G.}~\bibnamefont {Berghäuser}}, \bibinfo {author}
  {\bibfnamefont {J.}~\bibnamefont {Schaibley}}, \bibinfo {author}
  {\bibfnamefont {E.~J.}\ \bibnamefont {Seifert}}, \bibinfo {author}
  {\bibfnamefont {D.}~\bibnamefont {Pleskot}}, \bibinfo {author} {\bibfnamefont
  {N.~M.}\ \bibnamefont {Gabor}},\ and\ \bibinfo {author} {\bibnamefont
  {et~al.}},\ }\bibfield  {title} {\bibinfo {title} {Trion formation dynamics
  in monolayer transition metal dichalcogenides},\ }\bibfield  {journal}
  {\bibinfo  {journal} {Physical Review B}\ }\textbf {\bibinfo {volume} {93}},\
  \href {https://doi.org/10.1103/physrevb.93.041401}
  {10.1103/physrevb.93.041401} (\bibinfo {year} {2016})\BibitemShut {NoStop}%
\bibitem [{\citenamefont {Ye}\ \emph {et~al.}(2018)\citenamefont {Ye},
  \citenamefont {Yan}, \citenamefont {Niu}, \citenamefont {Li},\ and\
  \citenamefont {Zhang}}]{Dynamics_trions_free_carriers}%
  \BibitemOpen
  \bibfield  {author} {\bibinfo {author} {\bibfnamefont {J.}~\bibnamefont
  {Ye}}, \bibinfo {author} {\bibfnamefont {T.}~\bibnamefont {Yan}}, \bibinfo
  {author} {\bibfnamefont {B.}~\bibnamefont {Niu}}, \bibinfo {author}
  {\bibfnamefont {Y.}~\bibnamefont {Li}},\ and\ \bibinfo {author}
  {\bibfnamefont {X.}~\bibnamefont {Zhang}},\ }\bibfield  {title} {\bibinfo
  {title} {Nonlinear dynamics of trions under strong optical excitation in
  monolayer mose2},\ }\bibfield  {journal} {\bibinfo  {journal} {Scientific
  Reports}\ }\textbf {\bibinfo {volume} {8}},\ \href
  {https://doi.org/10.1038/s41598-018-20810-6} {10.1038/s41598-018-20810-6}
  (\bibinfo {year} {2018})\BibitemShut {NoStop}%
\bibitem [{\citenamefont {Yu}\ \emph {et~al.}(2016)\citenamefont {Yu},
  \citenamefont {Yu}, \citenamefont {Xu}, \citenamefont {Barrette},
  \citenamefont {Gundogdu},\ and\ \citenamefont
  {Cao}}]{Yu_Yu_Xu_Barrette_Gundogdu_Cao_2016}%
  \BibitemOpen
  \bibfield  {author} {\bibinfo {author} {\bibfnamefont {Y.}~\bibnamefont
  {Yu}}, \bibinfo {author} {\bibfnamefont {Y.}~\bibnamefont {Yu}}, \bibinfo
  {author} {\bibfnamefont {C.}~\bibnamefont {Xu}}, \bibinfo {author}
  {\bibfnamefont {A.}~\bibnamefont {Barrette}}, \bibinfo {author}
  {\bibfnamefont {K.}~\bibnamefont {Gundogdu}},\ and\ \bibinfo {author}
  {\bibfnamefont {L.}~\bibnamefont {Cao}},\ }\bibfield  {title} {\bibinfo
  {title} {Fundamental limits of exciton-exciton annihilation for light
  emission in transition metal dichalcogenide monolayers},\ }\bibfield
  {journal} {\bibinfo  {journal} {Physical Review B}\ }\textbf {\bibinfo
  {volume} {93}},\ \href {https://doi.org/10.1103/physrevb.93.201111}
  {10.1103/physrevb.93.201111} (\bibinfo {year} {2016})\BibitemShut {NoStop}%
\bibitem [{\citenamefont {Novoselov}\ \emph {et~al.}(2016)\citenamefont
  {Novoselov}, \citenamefont {Mishchenko}, \citenamefont {Carvalho},\ and\
  \citenamefont {Castro~Neto}}]{Novoselov_Mishchenko_Carvalho_Castro}%
  \BibitemOpen
  \bibfield  {author} {\bibinfo {author} {\bibfnamefont {K.~S.}\ \bibnamefont
  {Novoselov}}, \bibinfo {author} {\bibfnamefont {A.}~\bibnamefont
  {Mishchenko}}, \bibinfo {author} {\bibfnamefont {A.}~\bibnamefont
  {Carvalho}},\ and\ \bibinfo {author} {\bibfnamefont {A.~H.}\ \bibnamefont
  {Castro~Neto}},\ }\bibfield  {title} {\bibinfo {title} {2d materials and van
  der waals heterostructures},\ }\bibfield  {journal} {\bibinfo  {journal}
  {Science}\ }\textbf {\bibinfo {volume} {353}},\ \href
  {https://doi.org/10.1126/science.aac9439} {10.1126/science.aac9439} (\bibinfo
  {year} {2016})\BibitemShut {NoStop}%
\bibitem [{\citenamefont {Chen}\ \emph {et~al.}(2016)\citenamefont {Chen},
  \citenamefont {Wen}, \citenamefont {Zhang}, \citenamefont {Wu}, \citenamefont
  {Gong}, \citenamefont {Zhang}, \citenamefont {Yuan}, \citenamefont {Yi},
  \citenamefont {Lou}, \citenamefont {Ajayan},\ and\ \citenamefont
  {et~al.}}]{Inter_Ex_MoS2WS2}%
  \BibitemOpen
  \bibfield  {author} {\bibinfo {author} {\bibfnamefont {H.}~\bibnamefont
  {Chen}}, \bibinfo {author} {\bibfnamefont {X.}~\bibnamefont {Wen}}, \bibinfo
  {author} {\bibfnamefont {J.}~\bibnamefont {Zhang}}, \bibinfo {author}
  {\bibfnamefont {T.}~\bibnamefont {Wu}}, \bibinfo {author} {\bibfnamefont
  {Y.}~\bibnamefont {Gong}}, \bibinfo {author} {\bibfnamefont {X.}~\bibnamefont
  {Zhang}}, \bibinfo {author} {\bibfnamefont {J.}~\bibnamefont {Yuan}},
  \bibinfo {author} {\bibfnamefont {C.}~\bibnamefont {Yi}}, \bibinfo {author}
  {\bibfnamefont {J.}~\bibnamefont {Lou}}, \bibinfo {author} {\bibfnamefont
  {P.~M.}\ \bibnamefont {Ajayan}},\ and\ \bibinfo {author} {\bibnamefont
  {et~al.}},\ }\bibfield  {title} {\bibinfo {title} {Ultrafast formation of
  interlayer hot excitons in atomically thin mos2/ws2 heterostructures},\
  }\bibfield  {journal} {\bibinfo  {journal} {Nature Communications}\ }\textbf
  {\bibinfo {volume} {7}},\ \href {https://doi.org/10.1038/ncomms12512}
  {10.1038/ncomms12512} (\bibinfo {year} {2016})\BibitemShut {NoStop}%
\bibitem [{\citenamefont {Hanbicki}\ \emph {et~al.}(2018)\citenamefont
  {Hanbicki}, \citenamefont {Chuang}, \citenamefont {Rosenberger},
  \citenamefont {Hellberg}, \citenamefont {Sivaram}, \citenamefont {McCreary},
  \citenamefont {Mazin},\ and\ \citenamefont
  {Jonker}}]{Hanbicki_Chuang_Rosenberger_Hellberg_Sivaram_McCreary_Mazin_Jonker_2018}%
  \BibitemOpen
  \bibfield  {author} {\bibinfo {author} {\bibfnamefont {A.~T.}\ \bibnamefont
  {Hanbicki}}, \bibinfo {author} {\bibfnamefont {H.-J.}\ \bibnamefont
  {Chuang}}, \bibinfo {author} {\bibfnamefont {M.~R.}\ \bibnamefont
  {Rosenberger}}, \bibinfo {author} {\bibfnamefont {C.~S.}\ \bibnamefont
  {Hellberg}}, \bibinfo {author} {\bibfnamefont {S.~V.}\ \bibnamefont
  {Sivaram}}, \bibinfo {author} {\bibfnamefont {K.~M.}\ \bibnamefont
  {McCreary}}, \bibinfo {author} {\bibfnamefont {I.~I.}\ \bibnamefont
  {Mazin}},\ and\ \bibinfo {author} {\bibfnamefont {B.~T.}\ \bibnamefont
  {Jonker}},\ }\bibfield  {title} {\bibinfo {title} {Double indirect interlayer
  exciton in a mose2/wse2 van der waals heterostructure},\ }\href
  {https://doi.org/10.1021/acsnano.8b01369} {\bibfield  {journal} {\bibinfo
  {journal} {ACS Nano}\ }\textbf {\bibinfo {volume} {12}},\ \bibinfo {pages}
  {4719–4726} (\bibinfo {year} {2018})}\BibitemShut {NoStop}%
\bibitem [{\citenamefont {Zhu}\ \emph {et~al.}(2017)\citenamefont {Zhu},
  \citenamefont {Wang}, \citenamefont {Gong}, \citenamefont {Kim},
  \citenamefont {Hone},\ and\ \citenamefont
  {Zhu}}]{twisti_angle_dependece_interlayer_dynamics}%
  \BibitemOpen
  \bibfield  {author} {\bibinfo {author} {\bibfnamefont {H.}~\bibnamefont
  {Zhu}}, \bibinfo {author} {\bibfnamefont {J.}~\bibnamefont {Wang}}, \bibinfo
  {author} {\bibfnamefont {Z.}~\bibnamefont {Gong}}, \bibinfo {author}
  {\bibfnamefont {Y.~D.}\ \bibnamefont {Kim}}, \bibinfo {author} {\bibfnamefont
  {J.}~\bibnamefont {Hone}},\ and\ \bibinfo {author} {\bibfnamefont {X.-Y.}\
  \bibnamefont {Zhu}},\ }\bibfield  {title} {\bibinfo {title} {Interfacial
  charge transfer circumventing momentum mismatch at two-dimensional van der
  waals heterojunctions},\ }\href
  {https://doi.org/10.1021/acs.nanolett.7b00748} {\bibfield  {journal}
  {\bibinfo  {journal} {Nano Letters}\ }\textbf {\bibinfo {volume} {17}},\
  \bibinfo {pages} {3591–3598} (\bibinfo {year} {2017})}\BibitemShut
  {NoStop}%
\bibitem [{\citenamefont {Rivera}\ \emph {et~al.}(2016)\citenamefont {Rivera},
  \citenamefont {Seyler}, \citenamefont {Yu}, \citenamefont {Schaibley},
  \citenamefont {Yan}, \citenamefont {Mandrus}, \citenamefont {Yao},\ and\
  \citenamefont {Xu}}]{Rivera_field_tuning_interlayerEX}%
  \BibitemOpen
  \bibfield  {author} {\bibinfo {author} {\bibfnamefont {P.}~\bibnamefont
  {Rivera}}, \bibinfo {author} {\bibfnamefont {K.~L.}\ \bibnamefont {Seyler}},
  \bibinfo {author} {\bibfnamefont {H.}~\bibnamefont {Yu}}, \bibinfo {author}
  {\bibfnamefont {J.~R.}\ \bibnamefont {Schaibley}}, \bibinfo {author}
  {\bibfnamefont {J.}~\bibnamefont {Yan}}, \bibinfo {author} {\bibfnamefont
  {D.~G.}\ \bibnamefont {Mandrus}}, \bibinfo {author} {\bibfnamefont
  {W.}~\bibnamefont {Yao}},\ and\ \bibinfo {author} {\bibfnamefont
  {X.}~\bibnamefont {Xu}},\ }\bibfield  {title} {\bibinfo {title}
  {Valley-polarized exciton dynamics in a 2d semiconductor heterostructure},\
  }\href {https://doi.org/10.1126/science.aac7820} {\bibfield  {journal}
  {\bibinfo  {journal} {Science}\ }\textbf {\bibinfo {volume} {351}},\ \bibinfo
  {pages} {688–691} (\bibinfo {year} {2016})}\BibitemShut {NoStop}%
\bibitem [{\citenamefont {Jin}\ \emph {et~al.}(2018)\citenamefont {Jin},
  \citenamefont {Ma}, \citenamefont {Karni}, \citenamefont {Regan},
  \citenamefont {Wang},\ and\ \citenamefont {Heinz}}]{jin2018ultrafast}%
  \BibitemOpen
  \bibfield  {author} {\bibinfo {author} {\bibfnamefont {C.}~\bibnamefont
  {Jin}}, \bibinfo {author} {\bibfnamefont {E.~Y.}\ \bibnamefont {Ma}},
  \bibinfo {author} {\bibfnamefont {O.}~\bibnamefont {Karni}}, \bibinfo
  {author} {\bibfnamefont {E.~C.}\ \bibnamefont {Regan}}, \bibinfo {author}
  {\bibfnamefont {F.}~\bibnamefont {Wang}},\ and\ \bibinfo {author}
  {\bibfnamefont {T.~F.}\ \bibnamefont {Heinz}},\ }\bibfield  {title} {\bibinfo
  {title} {Ultrafast dynamics in van der waals heterostructures},\ }\href@noop
  {} {\bibfield  {journal} {\bibinfo  {journal} {Nature nanotechnology}\
  }\textbf {\bibinfo {volume} {13}},\ \bibinfo {pages} {994} (\bibinfo {year}
  {2018})}\BibitemShut {NoStop}%
\bibitem [{\citenamefont {Ma}\ \emph {et~al.}(2019)\citenamefont {Ma},
  \citenamefont {Guzelturk}, \citenamefont {Li}, \citenamefont {Cao},
  \citenamefont {Shen}, \citenamefont {Lindenberg},\ and\ \citenamefont
  {Heinz}}]{Ma_Guzelturk_Li_Cao_Shen_Lindenberg_Heinz_2019}%
  \BibitemOpen
  \bibfield  {author} {\bibinfo {author} {\bibfnamefont {E.~Y.}\ \bibnamefont
  {Ma}}, \bibinfo {author} {\bibfnamefont {B.}~\bibnamefont {Guzelturk}},
  \bibinfo {author} {\bibfnamefont {G.}~\bibnamefont {Li}}, \bibinfo {author}
  {\bibfnamefont {L.}~\bibnamefont {Cao}}, \bibinfo {author} {\bibfnamefont
  {Z.-X.}\ \bibnamefont {Shen}}, \bibinfo {author} {\bibfnamefont {A.~M.}\
  \bibnamefont {Lindenberg}},\ and\ \bibinfo {author} {\bibfnamefont {T.~F.}\
  \bibnamefont {Heinz}},\ }\bibfield  {title} {\bibinfo {title} {Recording
  interfacial currents on the subnanometer length and femtosecond time scale by
  terahertz emission},\ }\href {https://doi.org/10.1126/sciadv.aau0073}
  {\bibfield  {journal} {\bibinfo  {journal} {Science Advances}\ }\textbf
  {\bibinfo {volume} {5}},\ \bibinfo {pages} {eaau0073} (\bibinfo {year}
  {2019})}\BibitemShut {NoStop}%
\bibitem [{\citenamefont {Huang}\ \emph {et~al.}(2019)\citenamefont {Huang},
  \citenamefont {Yao}, \citenamefont {He}, \citenamefont {Zhu}, \citenamefont
  {Zhang}, \citenamefont {Bai},\ and\ \citenamefont
  {Xu}}]{Huang_Yao_He_Zhu_Zhang_Bai_Xu_2019}%
  \BibitemOpen
  \bibfield  {author} {\bibinfo {author} {\bibfnamefont {Y.}~\bibnamefont
  {Huang}}, \bibinfo {author} {\bibfnamefont {Z.}~\bibnamefont {Yao}}, \bibinfo
  {author} {\bibfnamefont {C.}~\bibnamefont {He}}, \bibinfo {author}
  {\bibfnamefont {L.}~\bibnamefont {Zhu}}, \bibinfo {author} {\bibfnamefont
  {L.}~\bibnamefont {Zhang}}, \bibinfo {author} {\bibfnamefont
  {J.}~\bibnamefont {Bai}},\ and\ \bibinfo {author} {\bibfnamefont
  {X.}~\bibnamefont {Xu}},\ }\bibfield  {title} {\bibinfo {title} {Terahertz
  surface and interface emission spectroscopy for advanced materials},\ }\href
  {https://doi.org/10.1088/1361-648x/ab00c0} {\bibfield  {journal} {\bibinfo
  {journal} {Journal of Physics: Condensed Matter}\ }\textbf {\bibinfo {volume}
  {31}},\ \bibinfo {pages} {153001} (\bibinfo {year} {2019})}\BibitemShut
  {NoStop}%
\bibitem [{\citenamefont {Stein}\ \emph {et~al.}(2020)\citenamefont {Stein},
  \citenamefont {Fuchs}, \citenamefont {Stolz}, \citenamefont {Mittleman},\
  and\ \citenamefont {Koch}}]{Stein_Fuchs_Stolz_Mittleman_Koch_2020}%
  \BibitemOpen
  \bibfield  {author} {\bibinfo {author} {\bibfnamefont {M.}~\bibnamefont
  {Stein}}, \bibinfo {author} {\bibfnamefont {C.}~\bibnamefont {Fuchs}},
  \bibinfo {author} {\bibfnamefont {W.}~\bibnamefont {Stolz}}, \bibinfo
  {author} {\bibfnamefont {D.~M.}\ \bibnamefont {Mittleman}},\ and\ \bibinfo
  {author} {\bibfnamefont {M.}~\bibnamefont {Koch}},\ }\bibfield  {title}
  {\bibinfo {title} {Direct probe of room-temperature quantum-tunneling
  processes in type-ii heterostructures using terahertz emission
  spectroscopy},\ }\bibfield  {journal} {\bibinfo  {journal} {Physical Review
  Applied}\ }\textbf {\bibinfo {volume} {13}},\ \href
  {https://doi.org/10.1103/physrevapplied.13.054073}
  {10.1103/physrevapplied.13.054073} (\bibinfo {year} {2020})\BibitemShut
  {NoStop}%
\bibitem [{\citenamefont {Seifert}\ \emph {et~al.}(2018)\citenamefont
  {Seifert}, \citenamefont {Jaiswal}, \citenamefont {Barker}, \citenamefont
  {Weber}, \citenamefont {Razdolski}, \citenamefont {Cramer}, \citenamefont
  {Gueckstock}, \citenamefont {Maehrlein}, \citenamefont {Nadvornik},
  \citenamefont {Watanabe},\ and\ \citenamefont
  {et~al.}}]{TobiasKa_resolution_Thz_spectr}%
  \BibitemOpen
  \bibfield  {author} {\bibinfo {author} {\bibfnamefont {T.~S.}\ \bibnamefont
  {Seifert}}, \bibinfo {author} {\bibfnamefont {S.}~\bibnamefont {Jaiswal}},
  \bibinfo {author} {\bibfnamefont {J.}~\bibnamefont {Barker}}, \bibinfo
  {author} {\bibfnamefont {S.~T.}\ \bibnamefont {Weber}}, \bibinfo {author}
  {\bibfnamefont {I.}~\bibnamefont {Razdolski}}, \bibinfo {author}
  {\bibfnamefont {J.}~\bibnamefont {Cramer}}, \bibinfo {author} {\bibfnamefont
  {O.}~\bibnamefont {Gueckstock}}, \bibinfo {author} {\bibfnamefont {S.~F.}\
  \bibnamefont {Maehrlein}}, \bibinfo {author} {\bibfnamefont {L.}~\bibnamefont
  {Nadvornik}}, \bibinfo {author} {\bibfnamefont {S.}~\bibnamefont
  {Watanabe}},\ and\ \bibinfo {author} {\bibnamefont {et~al.}},\ }\bibfield
  {title} {\bibinfo {title} {Femtosecond formation dynamics of the spin seebeck
  effect revealed by terahertz spectroscopy},\ }\bibfield  {journal} {\bibinfo
  {journal} {Nature Communications}\ }\textbf {\bibinfo {volume} {9}},\ \href
  {https://doi.org/10.1038/s41467-018-05135-2} {10.1038/s41467-018-05135-2}
  (\bibinfo {year} {2018})\BibitemShut {NoStop}%
\bibitem [{\citenamefont {Porer}\ \emph {et~al.}(2014)\citenamefont {Porer},
  \citenamefont {Ménard},\ and\ \citenamefont
  {Huber}}]{Record_sesnsetivity_EOS}%
  \BibitemOpen
  \bibfield  {author} {\bibinfo {author} {\bibfnamefont {M.}~\bibnamefont
  {Porer}}, \bibinfo {author} {\bibfnamefont {J.-M.}\ \bibnamefont {Ménard}},\
  and\ \bibinfo {author} {\bibfnamefont {R.}~\bibnamefont {Huber}},\ }\bibfield
   {title} {\bibinfo {title} {Shot noise reduced terahertz detection via
  spectrally postfiltered electro-optic sampling},\ }\href
  {https://doi.org/10.1364/ol.39.002435} {\bibfield  {journal} {\bibinfo
  {journal} {Optics Letters}\ }\textbf {\bibinfo {volume} {39}},\ \bibinfo
  {pages} {2435} (\bibinfo {year} {2014})}\BibitemShut {NoStop}%
\bibitem [{\citenamefont {Braun}\ \emph {et~al.}(2016)\citenamefont {Braun},
  \citenamefont {Mussler}, \citenamefont {Hruban}, \citenamefont
  {Konczykowski}, \citenamefont {Schumann}, \citenamefont {Wolf}, \citenamefont
  {Münzenberg}, \citenamefont {Perfetti},\ and\ \citenamefont
  {Kampfrath}}]{Braun_Mussler_Hruban_Konczykowski_Schumann_Wolf_Mnzenberg_Perfetti_Kampfrath_2016}%
  \BibitemOpen
  \bibfield  {author} {\bibinfo {author} {\bibfnamefont {L.}~\bibnamefont
  {Braun}}, \bibinfo {author} {\bibfnamefont {G.}~\bibnamefont {Mussler}},
  \bibinfo {author} {\bibfnamefont {A.}~\bibnamefont {Hruban}}, \bibinfo
  {author} {\bibfnamefont {M.}~\bibnamefont {Konczykowski}}, \bibinfo {author}
  {\bibfnamefont {T.}~\bibnamefont {Schumann}}, \bibinfo {author}
  {\bibfnamefont {M.}~\bibnamefont {Wolf}}, \bibinfo {author} {\bibfnamefont
  {M.}~\bibnamefont {Münzenberg}}, \bibinfo {author} {\bibfnamefont
  {L.}~\bibnamefont {Perfetti}},\ and\ \bibinfo {author} {\bibfnamefont
  {T.}~\bibnamefont {Kampfrath}},\ }\bibfield  {title} {\bibinfo {title}
  {Ultrafast photocurrents at the surface of the three-dimensional topological
  insulator bi2se3},\ }\bibfield  {journal} {\bibinfo  {journal} {Nature
  Communications}\ }\textbf {\bibinfo {volume} {7}},\ \href
  {https://doi.org/10.1038/ncomms13259} {10.1038/ncomms13259} (\bibinfo {year}
  {2016})\BibitemShut {NoStop}%
\bibitem [{\citenamefont {Zhang}\ \emph {et~al.}(1990)\citenamefont {Zhang},
  \citenamefont {Hu}, \citenamefont {Darrow},\ and\ \citenamefont
  {Auston}}]{Zhang_Hu_Darrow_Auston_1990}%
  \BibitemOpen
  \bibfield  {author} {\bibinfo {author} {\bibfnamefont {X.}~\bibnamefont
  {Zhang}}, \bibinfo {author} {\bibfnamefont {B.~B.}\ \bibnamefont {Hu}},
  \bibinfo {author} {\bibfnamefont {J.~T.}\ \bibnamefont {Darrow}},\ and\
  \bibinfo {author} {\bibfnamefont {D.~H.}\ \bibnamefont {Auston}},\ }\bibfield
   {title} {\bibinfo {title} {Generation of femtosecond electromagnetic pulses
  from semiconductor surfaces},\ }\bibfield  {journal} {\bibinfo  {journal}
  {Applied Physics Letters}\ }\textbf {\bibinfo {volume} {56}},\ \href
  {https://doi.org/10.1063/1.102601} {10.1063/1.102601} (\bibinfo {year}
  {1990})\BibitemShut {NoStop}%
\bibitem [{\citenamefont {Huang}\ \emph
  {et~al.}(2017{\natexlab{a}})\citenamefont {Huang}, \citenamefont {Zhu},
  \citenamefont {Yao}, \citenamefont {Zhang}, \citenamefont {He}, \citenamefont
  {Zhao}, \citenamefont {Bai},\ and\ \citenamefont {Xu}}]{huang2017terahertz}%
  \BibitemOpen
  \bibfield  {author} {\bibinfo {author} {\bibfnamefont {Y.}~\bibnamefont
  {Huang}}, \bibinfo {author} {\bibfnamefont {L.}~\bibnamefont {Zhu}}, \bibinfo
  {author} {\bibfnamefont {Z.}~\bibnamefont {Yao}}, \bibinfo {author}
  {\bibfnamefont {L.}~\bibnamefont {Zhang}}, \bibinfo {author} {\bibfnamefont
  {C.}~\bibnamefont {He}}, \bibinfo {author} {\bibfnamefont {Q.}~\bibnamefont
  {Zhao}}, \bibinfo {author} {\bibfnamefont {J.}~\bibnamefont {Bai}},\ and\
  \bibinfo {author} {\bibfnamefont {X.}~\bibnamefont {Xu}},\ }\bibfield
  {title} {\bibinfo {title} {Terahertz surface emission from layered mos2
  crystal: competition between surface optical rectification and surface
  photocurrent surge},\ }\href@noop {} {\bibfield  {journal} {\bibinfo
  {journal} {The Journal of Physical Chemistry C}\ }\textbf {\bibinfo {volume}
  {122}},\ \bibinfo {pages} {481} (\bibinfo {year}
  {2017}{\natexlab{a}})}\BibitemShut {NoStop}%
\bibitem [{\citenamefont {Zhang}\ \emph {et~al.}(2017)\citenamefont {Zhang},
  \citenamefont {Huang}, \citenamefont {Zhao}, \citenamefont {Zhu},
  \citenamefont {Yao}, \citenamefont {Zhou}, \citenamefont {Du},\ and\
  \citenamefont {Xu}}]{Zhang_Huang_Zhao_Zhu_Yao_Zhou_Du_Xu_2017}%
  \BibitemOpen
  \bibfield  {author} {\bibinfo {author} {\bibfnamefont {L.}~\bibnamefont
  {Zhang}}, \bibinfo {author} {\bibfnamefont {Y.}~\bibnamefont {Huang}},
  \bibinfo {author} {\bibfnamefont {Q.}~\bibnamefont {Zhao}}, \bibinfo {author}
  {\bibfnamefont {L.}~\bibnamefont {Zhu}}, \bibinfo {author} {\bibfnamefont
  {Z.}~\bibnamefont {Yao}}, \bibinfo {author} {\bibfnamefont {Y.}~\bibnamefont
  {Zhou}}, \bibinfo {author} {\bibfnamefont {W.}~\bibnamefont {Du}},\ and\
  \bibinfo {author} {\bibfnamefont {X.}~\bibnamefont {Xu}},\ }\bibfield
  {title} {\bibinfo {title} {Terahertz surface emission of d -band electrons
  from a layered tungsten disulfide crystal by the surface field},\ }\bibfield
  {journal} {\bibinfo  {journal} {Physical Review B}\ }\textbf {\bibinfo
  {volume} {96}},\ \href {https://doi.org/10.1103/physrevb.96.155202}
  {10.1103/physrevb.96.155202} (\bibinfo {year} {2017})\BibitemShut {NoStop}%
\bibitem [{\citenamefont {Si}\ \emph {et~al.}(2018)\citenamefont {Si},
  \citenamefont {Huang}, \citenamefont {Zhao}, \citenamefont {Zhu},
  \citenamefont {Zhang}, \citenamefont {Yao},\ and\ \citenamefont
  {Xu}}]{Si_Huang_Zhao_Zhu_Zhang_Yao_Xu_2018}%
  \BibitemOpen
  \bibfield  {author} {\bibinfo {author} {\bibfnamefont {K.}~\bibnamefont
  {Si}}, \bibinfo {author} {\bibfnamefont {Y.}~\bibnamefont {Huang}}, \bibinfo
  {author} {\bibfnamefont {Q.}~\bibnamefont {Zhao}}, \bibinfo {author}
  {\bibfnamefont {L.}~\bibnamefont {Zhu}}, \bibinfo {author} {\bibfnamefont
  {L.}~\bibnamefont {Zhang}}, \bibinfo {author} {\bibfnamefont
  {Z.}~\bibnamefont {Yao}},\ and\ \bibinfo {author} {\bibfnamefont
  {X.}~\bibnamefont {Xu}},\ }\bibfield  {title} {\bibinfo {title} {Terahertz
  surface emission from layered semiconductor wse2},\ }\bibfield  {journal}
  {\bibinfo  {journal} {Applied Surface Science}\ }\textbf {\bibinfo {volume}
  {448}},\ \href {https://doi.org/10.1016/j.apsusc.2018.04.117}
  {10.1016/j.apsusc.2018.04.117} (\bibinfo {year} {2018})\BibitemShut {NoStop}%
\bibitem [{\citenamefont {Zhang}\ \emph {et~al.}(1992)\citenamefont {Zhang},
  \citenamefont {Jin}, \citenamefont {Yang},\ and\ \citenamefont
  {Schowalter}}]{Zhang_Jin_Yang_Schowalter_1992}%
  \BibitemOpen
  \bibfield  {author} {\bibinfo {author} {\bibfnamefont {X.-C.}\ \bibnamefont
  {Zhang}}, \bibinfo {author} {\bibfnamefont {Y.}~\bibnamefont {Jin}}, \bibinfo
  {author} {\bibfnamefont {K.}~\bibnamefont {Yang}},\ and\ \bibinfo {author}
  {\bibfnamefont {L.~J.}\ \bibnamefont {Schowalter}},\ }\bibfield  {title}
  {\bibinfo {title} {Resonant nonlinear susceptibility near the gaas band
  gap},\ }\href {https://doi.org/10.1103/physrevlett.69.2303} {\bibfield
  {journal} {\bibinfo  {journal} {Physical Review Letters}\ }\textbf {\bibinfo
  {volume} {69}},\ \bibinfo {pages} {2303–2306} (\bibinfo {year}
  {1992})}\BibitemShut {NoStop}%
\bibitem [{\citenamefont {Côté}\ \emph {et~al.}(2002)\citenamefont {Côté},
  \citenamefont {Laman},\ and\ \citenamefont {van
  Driel}}]{Cote_Laman_vanDriel_2002}%
  \BibitemOpen
  \bibfield  {author} {\bibinfo {author} {\bibfnamefont {D.}~\bibnamefont
  {Côté}}, \bibinfo {author} {\bibfnamefont {N.}~\bibnamefont {Laman}},\ and\
  \bibinfo {author} {\bibfnamefont {H.~M.}\ \bibnamefont {van Driel}},\
  }\bibfield  {title} {\bibinfo {title} {Rectification and shift currents in
  gaas},\ }\href {https://doi.org/10.1063/1.1436530} {\bibfield  {journal}
  {\bibinfo  {journal} {Applied Physics Letters}\ }\textbf {\bibinfo {volume}
  {80}},\ \bibinfo {pages} {905–907} (\bibinfo {year} {2002})}\BibitemShut
  {NoStop}%
\bibitem [{\citenamefont {Sipe}\ and\ \citenamefont
  {Shkrebtii}(2000)}]{Sipe_Shkrebtii_2000}%
  \BibitemOpen
  \bibfield  {author} {\bibinfo {author} {\bibfnamefont {J.~E.}\ \bibnamefont
  {Sipe}}\ and\ \bibinfo {author} {\bibfnamefont {A.~I.}\ \bibnamefont
  {Shkrebtii}},\ }\bibfield  {title} {\bibinfo {title} {Second-order optical
  response in semiconductors},\ }\href
  {https://doi.org/10.1103/physrevb.61.5337} {\bibfield  {journal} {\bibinfo
  {journal} {Physical Review B}\ }\textbf {\bibinfo {volume} {61}},\ \bibinfo
  {pages} {5337–5352} (\bibinfo {year} {2000})}\BibitemShut {NoStop}%
\bibitem [{\citenamefont {Nastos}\ and\ \citenamefont
  {Sipe}(2006)}]{Nastos_Sipe_2006}%
  \BibitemOpen
  \bibfield  {author} {\bibinfo {author} {\bibfnamefont {F.}~\bibnamefont
  {Nastos}}\ and\ \bibinfo {author} {\bibfnamefont {J.~E.}\ \bibnamefont
  {Sipe}},\ }\bibfield  {title} {\bibinfo {title} {Optical rectification and
  shift currents in gaas and gap response: Below and above the band gap},\
  }\bibfield  {journal} {\bibinfo  {journal} {Physical Review B}\ }\textbf
  {\bibinfo {volume} {74}},\ \href {https://doi.org/10.1103/physrevb.74.035201}
  {10.1103/physrevb.74.035201} (\bibinfo {year} {2006})\BibitemShut {NoStop}%
\bibitem [{\citenamefont {Nastos}\ and\ \citenamefont
  {Sipe}(2010)}]{nastos2010optical}%
  \BibitemOpen
  \bibfield  {author} {\bibinfo {author} {\bibfnamefont {F.}~\bibnamefont
  {Nastos}}\ and\ \bibinfo {author} {\bibfnamefont {J.}~\bibnamefont {Sipe}},\
  }\bibfield  {title} {\bibinfo {title} {Optical rectification and current
  injection in unbiased semiconductors},\ }\href@noop {} {\bibfield  {journal}
  {\bibinfo  {journal} {Physical Review B}\ }\textbf {\bibinfo {volume} {82}},\
  \bibinfo {pages} {235204} (\bibinfo {year} {2010})}\BibitemShut {NoStop}%
\bibitem [{\citenamefont {Göbel}\ \emph {et~al.}(1990)\citenamefont {Göbel},
  \citenamefont {Leo}, \citenamefont {Damen}, \citenamefont {Shah},
  \citenamefont {Schmitt-Rink}, \citenamefont {Schäfer}, \citenamefont
  {Müller},\ and\ \citenamefont
  {Köhler}}]{Gobel_Leo_Damen_Shah_Schmitt-Rink_Schafer_Muller_Kohler_1990}%
  \BibitemOpen
  \bibfield  {author} {\bibinfo {author} {\bibfnamefont {E.~O.}\ \bibnamefont
  {Göbel}}, \bibinfo {author} {\bibfnamefont {K.}~\bibnamefont {Leo}},
  \bibinfo {author} {\bibfnamefont {T.~C.}\ \bibnamefont {Damen}}, \bibinfo
  {author} {\bibfnamefont {J.}~\bibnamefont {Shah}}, \bibinfo {author}
  {\bibfnamefont {S.}~\bibnamefont {Schmitt-Rink}}, \bibinfo {author}
  {\bibfnamefont {W.}~\bibnamefont {Schäfer}}, \bibinfo {author}
  {\bibfnamefont {J.~F.}\ \bibnamefont {Müller}},\ and\ \bibinfo {author}
  {\bibfnamefont {K.}~\bibnamefont {Köhler}},\ }\bibfield  {title} {\bibinfo
  {title} {Quantum beats of excitons in quantum wells},\ }\href
  {https://doi.org/10.1103/physrevlett.64.1801} {\bibfield  {journal} {\bibinfo
   {journal} {Physical Review Letters}\ }\textbf {\bibinfo {volume} {64}},\
  \bibinfo {pages} {1801–1804} (\bibinfo {year} {1990})}\BibitemShut
  {NoStop}%
\bibitem [{\citenamefont {Planken}\ \emph {et~al.}(1992)\citenamefont
  {Planken}, \citenamefont {Nuss}, \citenamefont {Brener}, \citenamefont
  {Goossen}, \citenamefont {Luo}, \citenamefont {Chuang},\ and\ \citenamefont
  {Pfeiffer}}]{coherent_osc_in_gaas}%
  \BibitemOpen
  \bibfield  {author} {\bibinfo {author} {\bibfnamefont {P.~C.}\ \bibnamefont
  {Planken}}, \bibinfo {author} {\bibfnamefont {M.~C.}\ \bibnamefont {Nuss}},
  \bibinfo {author} {\bibfnamefont {I.}~\bibnamefont {Brener}}, \bibinfo
  {author} {\bibfnamefont {K.~W.}\ \bibnamefont {Goossen}}, \bibinfo {author}
  {\bibfnamefont {M.~S.}\ \bibnamefont {Luo}}, \bibinfo {author} {\bibfnamefont
  {S.~L.}\ \bibnamefont {Chuang}},\ and\ \bibinfo {author} {\bibfnamefont
  {L.}~\bibnamefont {Pfeiffer}},\ }\bibfield  {title} {\bibinfo {title}
  {Terahertz emission in single quantum wells after coherent optical excitation
  of light hole and heavy hole excitons},\ }\href@noop {} {\bibfield  {journal}
  {\bibinfo  {journal} {Physical review letters}\ }\textbf {\bibinfo {volume}
  {69}},\ \bibinfo {pages} {3800} (\bibinfo {year} {1992})}\BibitemShut
  {NoStop}%
\bibitem [{\citenamefont {Brener}\ \emph {et~al.}(1994)\citenamefont {Brener},
  \citenamefont {Planken}, \citenamefont {Nuss}, \citenamefont {Luo},
  \citenamefont {Chuang}, \citenamefont {Pfeiffer}, \citenamefont {Leaird},\
  and\ \citenamefont
  {Weiner}}]{Brener_Planken_Nuss_Luo_Chuang_Pfeiffer_Leaird_Weiner_1994}%
  \BibitemOpen
  \bibfield  {author} {\bibinfo {author} {\bibfnamefont {I.}~\bibnamefont
  {Brener}}, \bibinfo {author} {\bibfnamefont {P.~C.~M.}\ \bibnamefont
  {Planken}}, \bibinfo {author} {\bibfnamefont {M.~C.}\ \bibnamefont {Nuss}},
  \bibinfo {author} {\bibfnamefont {M.~S.~C.}\ \bibnamefont {Luo}}, \bibinfo
  {author} {\bibfnamefont {S.~L.}\ \bibnamefont {Chuang}}, \bibinfo {author}
  {\bibfnamefont {L.}~\bibnamefont {Pfeiffer}}, \bibinfo {author}
  {\bibfnamefont {D.~E.}\ \bibnamefont {Leaird}},\ and\ \bibinfo {author}
  {\bibfnamefont {A.~M.}\ \bibnamefont {Weiner}},\ }\bibfield  {title}
  {\bibinfo {title} {Coherent control of terahertz emission and carrier
  populations in semiconductor heterostructures},\ }\href
  {https://doi.org/10.1364/josab.11.002457} {\bibfield  {journal} {\bibinfo
  {journal} {Journal of the Optical Society of America B}\ }\textbf {\bibinfo
  {volume} {11}},\ \bibinfo {pages} {2457} (\bibinfo {year}
  {1994})}\BibitemShut {NoStop}%
\bibitem [{\citenamefont {Cadiz}\ \emph {et~al.}(2017)\citenamefont {Cadiz},
  \citenamefont {Courtade}, \citenamefont {Robert}, \citenamefont {Wang},
  \citenamefont {Shen}, \citenamefont {Cai}, \citenamefont {Taniguchi},
  \citenamefont {Watanabe}, \citenamefont {Carrere}, \citenamefont {Lagarde},\
  and\ \citenamefont
  {et~al.}}]{Cadiz_Courtade_Robert_Wang_Shen_Cai_Taniguchi_Watanabe_Carrere_Lagar}%
  \BibitemOpen
  \bibfield  {author} {\bibinfo {author} {\bibfnamefont {F.}~\bibnamefont
  {Cadiz}}, \bibinfo {author} {\bibfnamefont {E.}~\bibnamefont {Courtade}},
  \bibinfo {author} {\bibfnamefont {C.}~\bibnamefont {Robert}}, \bibinfo
  {author} {\bibfnamefont {G.}~\bibnamefont {Wang}}, \bibinfo {author}
  {\bibfnamefont {Y.}~\bibnamefont {Shen}}, \bibinfo {author} {\bibfnamefont
  {H.}~\bibnamefont {Cai}}, \bibinfo {author} {\bibfnamefont {T.}~\bibnamefont
  {Taniguchi}}, \bibinfo {author} {\bibfnamefont {K.}~\bibnamefont {Watanabe}},
  \bibinfo {author} {\bibfnamefont {H.}~\bibnamefont {Carrere}}, \bibinfo
  {author} {\bibfnamefont {D.}~\bibnamefont {Lagarde}},\ and\ \bibinfo {author}
  {\bibnamefont {et~al.}},\ }\bibfield  {title} {\bibinfo {title} {Excitonic
  linewidth approaching the homogeneous limit in mos2 -based van der waals
  heterostructures},\ }\bibfield  {journal} {\bibinfo  {journal} {Physical
  Review X}\ }\textbf {\bibinfo {volume} {7}},\ \href
  {https://doi.org/10.1103/physrevx.7.021026} {10.1103/physrevx.7.021026}
  (\bibinfo {year} {2017})\BibitemShut {NoStop}%
\bibitem [{\citenamefont {Tonndorf}\ \emph {et~al.}(2013)\citenamefont
  {Tonndorf}, \citenamefont {Schmidt}, \citenamefont {Böttger}, \citenamefont
  {Zhang}, \citenamefont {Börner}, \citenamefont {Liebig}, \citenamefont
  {Albrecht}, \citenamefont {Kloc}, \citenamefont {Gordan}, \citenamefont
  {Zahn},\ and\ \citenamefont {et~al.}}]{About_PLandRaman_of_all_TMDCs}%
  \BibitemOpen
  \bibfield  {author} {\bibinfo {author} {\bibfnamefont {P.}~\bibnamefont
  {Tonndorf}}, \bibinfo {author} {\bibfnamefont {R.}~\bibnamefont {Schmidt}},
  \bibinfo {author} {\bibfnamefont {P.}~\bibnamefont {Böttger}}, \bibinfo
  {author} {\bibfnamefont {X.}~\bibnamefont {Zhang}}, \bibinfo {author}
  {\bibfnamefont {J.}~\bibnamefont {Börner}}, \bibinfo {author} {\bibfnamefont
  {A.}~\bibnamefont {Liebig}}, \bibinfo {author} {\bibfnamefont
  {M.}~\bibnamefont {Albrecht}}, \bibinfo {author} {\bibfnamefont
  {C.}~\bibnamefont {Kloc}}, \bibinfo {author} {\bibfnamefont {O.}~\bibnamefont
  {Gordan}}, \bibinfo {author} {\bibfnamefont {D.~R.~T.}\ \bibnamefont
  {Zahn}},\ and\ \bibinfo {author} {\bibnamefont {et~al.}},\ }\bibfield
  {title} {\bibinfo {title} {Photoluminescence emission and raman response of
  monolayer mos$_2$, mose$_2$, and wse$_2$},\ }\href
  {https://doi.org/10.1364/oe.21.004908} {\bibfield  {journal} {\bibinfo
  {journal} {Optics Express}\ }\textbf {\bibinfo {volume} {21}},\ \bibinfo
  {pages} {4908} (\bibinfo {year} {2013})}\BibitemShut {NoStop}%
\bibitem [{\citenamefont {Seifert}(2017)}]{phdthesisTHz}%
  \BibitemOpen
  \bibfield  {author} {\bibinfo {author} {\bibfnamefont {T.~S.}\ \bibnamefont
  {Seifert}},\ }\emph {\bibinfo {title} {Spintronics with Terahertz Radiation:
  Probing and driving spins at highest frequencies}},\ \href@noop {} {Ph.D.
  thesis},\ \bibinfo  {school} {Freie Universität Berlin}, \bibinfo {address}
  {Berlin} (\bibinfo {year} {2017})\BibitemShut {NoStop}%
\bibitem [{\citenamefont {Kampfrath}\ \emph {et~al.}(2007)\citenamefont
  {Kampfrath}, \citenamefont {Nötzold},\ and\ \citenamefont
  {Wolf}}]{Kampfrath_EOS}%
  \BibitemOpen
  \bibfield  {author} {\bibinfo {author} {\bibfnamefont {T.}~\bibnamefont
  {Kampfrath}}, \bibinfo {author} {\bibfnamefont {J.}~\bibnamefont
  {Nötzold}},\ and\ \bibinfo {author} {\bibfnamefont {M.}~\bibnamefont
  {Wolf}},\ }\bibfield  {title} {\bibinfo {title} {Sampling of broadband
  terahertz pulses with thick electro-optic crystals},\ }\href
  {https://doi.org/10.1063/1.2746939} {\bibfield  {journal} {\bibinfo
  {journal} {Applied Physics Letters}\ }\textbf {\bibinfo {volume} {90}},\
  \bibinfo {pages} {231113} (\bibinfo {year} {2007})}\BibitemShut {NoStop}%
\bibitem [{\citenamefont {Shimada}\ \emph {et~al.}(2014)\citenamefont
  {Shimada}, \citenamefont {Kamaraju}, \citenamefont {Frischkorn},
  \citenamefont {Wolf},\ and\ \citenamefont
  {Kampfrath}}]{Shimada_Kamaraju_Frischkorn_Wolf_Kampfrath_2014}%
  \BibitemOpen
  \bibfield  {author} {\bibinfo {author} {\bibfnamefont {T.}~\bibnamefont
  {Shimada}}, \bibinfo {author} {\bibfnamefont {N.}~\bibnamefont {Kamaraju}},
  \bibinfo {author} {\bibfnamefont {C.}~\bibnamefont {Frischkorn}}, \bibinfo
  {author} {\bibfnamefont {M.}~\bibnamefont {Wolf}},\ and\ \bibinfo {author}
  {\bibfnamefont {T.}~\bibnamefont {Kampfrath}},\ }\bibfield  {title} {\bibinfo
  {title} {Indication of te segregation in laser-irradiated znte observed by in
  situ coherent-phonon spectroscopy},\ }\href
  {https://doi.org/10.1063/1.4896039} {\bibfield  {journal} {\bibinfo
  {journal} {Applied Physics Letters}\ }\textbf {\bibinfo {volume} {105}},\
  \bibinfo {pages} {111908} (\bibinfo {year} {2014})}\BibitemShut {NoStop}%
\bibitem [{\citenamefont {Zhao}\ \emph {et~al.}(2010)\citenamefont {Zhao},
  \citenamefont {Schwagmann}, \citenamefont {Ospald}, \citenamefont {Driscoll},
  \citenamefont {Lu}, \citenamefont {Gossard},\ and\ \citenamefont
  {Smet}}]{Zhao_Schwagmann_Ospald_Driscoll_Lu_Gossard_Smet_2010}%
  \BibitemOpen
  \bibfield  {author} {\bibinfo {author} {\bibfnamefont {Z.}~\bibnamefont
  {Zhao}}, \bibinfo {author} {\bibfnamefont {A.}~\bibnamefont {Schwagmann}},
  \bibinfo {author} {\bibfnamefont {F.}~\bibnamefont {Ospald}}, \bibinfo
  {author} {\bibfnamefont {D.~C.}\ \bibnamefont {Driscoll}}, \bibinfo {author}
  {\bibfnamefont {H.}~\bibnamefont {Lu}}, \bibinfo {author} {\bibfnamefont
  {A.~C.}\ \bibnamefont {Gossard}},\ and\ \bibinfo {author} {\bibfnamefont
  {J.~H.}\ \bibnamefont {Smet}},\ }\bibfield  {title} {\bibinfo {title}
  {Thickness dependence of the terahertz response in 〈110〉-oriented gaas
  crystals for electro-optic sampling at 155 µm},\ }\href
  {https://doi.org/10.1364/oe.18.015956} {\bibfield  {journal} {\bibinfo
  {journal} {Optics Express}\ }\textbf {\bibinfo {volume} {18}},\ \bibinfo
  {pages} {15956} (\bibinfo {year} {2010})}\BibitemShut {NoStop}%
\bibitem [{\citenamefont {Beck}\ \emph {et~al.}(2010)\citenamefont {Beck},
  \citenamefont {Schäfer}, \citenamefont {Klatt}, \citenamefont {Demsar},
  \citenamefont {Winnerl}, \citenamefont {Helm},\ and\ \citenamefont
  {Dekorsy}}]{Beck_Schfer_Klatt_Demsar_Winnerl_Helm_Dekorsy_2010}%
  \BibitemOpen
  \bibfield  {author} {\bibinfo {author} {\bibfnamefont {M.}~\bibnamefont
  {Beck}}, \bibinfo {author} {\bibfnamefont {H.}~\bibnamefont {Schäfer}},
  \bibinfo {author} {\bibfnamefont {G.}~\bibnamefont {Klatt}}, \bibinfo
  {author} {\bibfnamefont {J.}~\bibnamefont {Demsar}}, \bibinfo {author}
  {\bibfnamefont {S.}~\bibnamefont {Winnerl}}, \bibinfo {author} {\bibfnamefont
  {M.}~\bibnamefont {Helm}},\ and\ \bibinfo {author} {\bibfnamefont
  {T.}~\bibnamefont {Dekorsy}},\ }\bibfield  {title} {\bibinfo {title}
  {Impulsive terahertz radiation with high electric fields from an
  amplifier-driven large-area photoconductive antenna},\ }\href
  {https://doi.org/10.1364/oe.18.009251} {\bibfield  {journal} {\bibinfo
  {journal} {Optics Express}\ }\textbf {\bibinfo {volume} {18}},\ \bibinfo
  {pages} {9251} (\bibinfo {year} {2010})}\BibitemShut {NoStop}%
\bibitem [{\citenamefont {Johnston}\ \emph {et~al.}(2002)\citenamefont
  {Johnston}, \citenamefont {Whittaker}, \citenamefont {Corchia}, \citenamefont
  {Davies},\ and\ \citenamefont
  {Linfield}}]{Johnston_Whittaker_Corchia_Davies_Linfield_2002}%
  \BibitemOpen
  \bibfield  {author} {\bibinfo {author} {\bibfnamefont {M.~B.}\ \bibnamefont
  {Johnston}}, \bibinfo {author} {\bibfnamefont {D.~M.}\ \bibnamefont
  {Whittaker}}, \bibinfo {author} {\bibfnamefont {A.}~\bibnamefont {Corchia}},
  \bibinfo {author} {\bibfnamefont {A.~G.}\ \bibnamefont {Davies}},\ and\
  \bibinfo {author} {\bibfnamefont {E.~H.}\ \bibnamefont {Linfield}},\
  }\bibfield  {title} {\bibinfo {title} {Simulation of terahertz generation at
  semiconductor surfaces},\ }\bibfield  {journal} {\bibinfo  {journal}
  {Physical Review B}\ }\textbf {\bibinfo {volume} {65}},\ \href
  {https://doi.org/10.1103/physrevb.65.165301} {10.1103/physrevb.65.165301}
  (\bibinfo {year} {2002})\BibitemShut {NoStop}%
\bibitem [{\citenamefont {Heyman}\ \emph {et~al.}(2003)\citenamefont {Heyman},
  \citenamefont {Coates}, \citenamefont {Reinhardt},\ and\ \citenamefont
  {Strasser}}]{Heyman_Coates_Reinhardt_Strasser_2003}%
  \BibitemOpen
  \bibfield  {author} {\bibinfo {author} {\bibfnamefont {J.~N.}\ \bibnamefont
  {Heyman}}, \bibinfo {author} {\bibfnamefont {N.}~\bibnamefont {Coates}},
  \bibinfo {author} {\bibfnamefont {A.}~\bibnamefont {Reinhardt}},\ and\
  \bibinfo {author} {\bibfnamefont {G.}~\bibnamefont {Strasser}},\ }\bibfield
  {title} {\bibinfo {title} {Diffusion and drift in terahertz emission at gaas
  surfaces},\ }\href {https://doi.org/10.1063/1.1636821} {\bibfield  {journal}
  {\bibinfo  {journal} {Applied Physics Letters}\ }\textbf {\bibinfo {volume}
  {83}},\ \bibinfo {pages} {5476–5478} (\bibinfo {year} {2003})}\BibitemShut
  {NoStop}%
\bibitem [{\citenamefont {Fregoso}(2019)}]{Fregoso_2019}%
  \BibitemOpen
  \bibfield  {author} {\bibinfo {author} {\bibfnamefont {B.~M.}\ \bibnamefont
  {Fregoso}},\ }\bibfield  {title} {\bibinfo {title} {Bulk photovoltaic effects
  in the presence of a static electric field},\ }\bibfield  {journal} {\bibinfo
   {journal} {Physical Review B}\ }\textbf {\bibinfo {volume} {100}},\ \href
  {https://doi.org/10.1103/physrevb.100.064301} {10.1103/physrevb.100.064301}
  (\bibinfo {year} {2019})\BibitemShut {NoStop}%
\bibitem [{\citenamefont {Manzeli}\ \emph {et~al.}(2017)\citenamefont
  {Manzeli}, \citenamefont {Ovchinnikov}, \citenamefont {Pasquier},
  \citenamefont {Yazyev},\ and\ \citenamefont
  {Kis}}]{Manzeli_Ovchinnikov_Pasquier_Yazyev_Kis_2017}%
  \BibitemOpen
  \bibfield  {author} {\bibinfo {author} {\bibfnamefont {S.}~\bibnamefont
  {Manzeli}}, \bibinfo {author} {\bibfnamefont {D.}~\bibnamefont
  {Ovchinnikov}}, \bibinfo {author} {\bibfnamefont {D.}~\bibnamefont
  {Pasquier}}, \bibinfo {author} {\bibfnamefont {O.~V.}\ \bibnamefont
  {Yazyev}},\ and\ \bibinfo {author} {\bibfnamefont {A.}~\bibnamefont {Kis}},\
  }\bibfield  {title} {\bibinfo {title} {2d transition metal dichalcogenides},\
  }\bibfield  {journal} {\bibinfo  {journal} {Nature Reviews Materials}\
  }\textbf {\bibinfo {volume} {2}},\ \href
  {https://doi.org/10.1038/natrevmats.2017.33} {10.1038/natrevmats.2017.33}
  (\bibinfo {year} {2017})\BibitemShut {NoStop}%
\bibitem [{\citenamefont {Steinberg}\ \emph {et~al.}(2010)\citenamefont
  {Steinberg}, \citenamefont {Gardner}, \citenamefont {Lee},\ and\
  \citenamefont
  {Jarillo-Herrero}}]{Steinberg_Gardner_Lee_Jarillo-Herrero_2010}%
  \BibitemOpen
  \bibfield  {author} {\bibinfo {author} {\bibfnamefont {H.}~\bibnamefont
  {Steinberg}}, \bibinfo {author} {\bibfnamefont {D.~R.}\ \bibnamefont
  {Gardner}}, \bibinfo {author} {\bibfnamefont {Y.~S.}\ \bibnamefont {Lee}},\
  and\ \bibinfo {author} {\bibfnamefont {P.}~\bibnamefont {Jarillo-Herrero}},\
  }\bibfield  {title} {\bibinfo {title} {Surface state transport and ambipolar
  electric field effect in bi2se3 nanodevices},\ }\href
  {https://doi.org/10.1021/nl1032183} {\bibfield  {journal} {\bibinfo
  {journal} {Nano Letters}\ }\textbf {\bibinfo {volume} {10}},\ \bibinfo
  {pages} {5032–5036} (\bibinfo {year} {2010})}\BibitemShut {NoStop}%
\bibitem [{\citenamefont {McIver}\ \emph {et~al.}(2012)\citenamefont {McIver},
  \citenamefont {Hsieh}, \citenamefont {Drapcho}, \citenamefont {Torchinsky},
  \citenamefont {Gardner}, \citenamefont {Lee},\ and\ \citenamefont
  {Gedik}}]{McIver_Hsieh_Drapcho_Torchinsky_Gardner_Lee_Gedik_2012}%
  \BibitemOpen
  \bibfield  {author} {\bibinfo {author} {\bibfnamefont {J.~W.}\ \bibnamefont
  {McIver}}, \bibinfo {author} {\bibfnamefont {D.}~\bibnamefont {Hsieh}},
  \bibinfo {author} {\bibfnamefont {S.~G.}\ \bibnamefont {Drapcho}}, \bibinfo
  {author} {\bibfnamefont {D.~H.}\ \bibnamefont {Torchinsky}}, \bibinfo
  {author} {\bibfnamefont {D.~R.}\ \bibnamefont {Gardner}}, \bibinfo {author}
  {\bibfnamefont {Y.~S.}\ \bibnamefont {Lee}},\ and\ \bibinfo {author}
  {\bibfnamefont {N.}~\bibnamefont {Gedik}},\ }\bibfield  {title} {\bibinfo
  {title} {Theoretical and experimental study of second harmonic generation
  from the surface of the topological insulator bi2se3},\ }\bibfield  {journal}
  {\bibinfo  {journal} {Physical Review B}\ }\textbf {\bibinfo {volume} {86}},\
  \href {https://doi.org/10.1103/physrevb.86.035327}
  {10.1103/physrevb.86.035327} (\bibinfo {year} {2012})\BibitemShut {NoStop}%
\bibitem [{\citenamefont {Huang}\ \emph
  {et~al.}(2017{\natexlab{b}})\citenamefont {Huang}, \citenamefont {Zhu},
  \citenamefont {Zhao}, \citenamefont {Guo}, \citenamefont {Ren}, \citenamefont
  {Bai},\ and\ \citenamefont {Xu}}]{Huang_Zhu_Zhao_Guo_Ren_Bai_Xu_2017}%
  \BibitemOpen
  \bibfield  {author} {\bibinfo {author} {\bibfnamefont {Y.}~\bibnamefont
  {Huang}}, \bibinfo {author} {\bibfnamefont {L.}~\bibnamefont {Zhu}}, \bibinfo
  {author} {\bibfnamefont {Q.}~\bibnamefont {Zhao}}, \bibinfo {author}
  {\bibfnamefont {Y.}~\bibnamefont {Guo}}, \bibinfo {author} {\bibfnamefont
  {Z.}~\bibnamefont {Ren}}, \bibinfo {author} {\bibfnamefont {J.}~\bibnamefont
  {Bai}},\ and\ \bibinfo {author} {\bibfnamefont {X.}~\bibnamefont {Xu}},\
  }\bibfield  {title} {\bibinfo {title} {Surface optical rectification from
  layered mos2 crystal by thz time-domain surface emission spectroscopy},\
  }\href {https://doi.org/10.1021/acsami.6b13961} {\bibfield  {journal}
  {\bibinfo  {journal} {ACS Applied Materials and Interfaces}\ }\textbf {\bibinfo
  {volume} {9}},\ \bibinfo {pages} {4956–4965} (\bibinfo {year}
  {2017}{\natexlab{b}})}\BibitemShut {NoStop}%
\bibitem [{\citenamefont {Ferguson}\ and\ \citenamefont
  {Zhang}(2002)}]{Ferguson_Zhang_2002}%
  \BibitemOpen
  \bibfield  {author} {\bibinfo {author} {\bibfnamefont {B.}~\bibnamefont
  {Ferguson}}\ and\ \bibinfo {author} {\bibfnamefont {X.-C.}\ \bibnamefont
  {Zhang}},\ }\bibfield  {title} {\bibinfo {title} {Materials for terahertz
  science and technology},\ }\href {https://doi.org/10.1038/nmat708} {\bibfield
   {journal} {\bibinfo  {journal} {Nature Materials}\ }\textbf {\bibinfo
  {volume} {1}},\ \bibinfo {pages} {26–33} (\bibinfo {year}
  {2002})}\BibitemShut {NoStop}%
\bibitem [{\citenamefont {Kyriienko}\ \emph {et~al.}(2013)\citenamefont
  {Kyriienko}, \citenamefont {Kavokin},\ and\ \citenamefont
  {Shelykh}}]{kyriienko2013superradiant}%
  \BibitemOpen
  \bibfield  {author} {\bibinfo {author} {\bibfnamefont {O.}~\bibnamefont
  {Kyriienko}}, \bibinfo {author} {\bibfnamefont {A.}~\bibnamefont {Kavokin}},\
  and\ \bibinfo {author} {\bibfnamefont {I.}~\bibnamefont {Shelykh}},\
  }\bibfield  {title} {\bibinfo {title} {Superradiant terahertz emission by
  dipolaritons},\ }\href@noop {} {\bibfield  {journal} {\bibinfo  {journal}
  {Physical review letters}\ }\textbf {\bibinfo {volume} {111}},\ \bibinfo
  {pages} {176401} (\bibinfo {year} {2013})}\BibitemShut {NoStop}%
\bibitem [{\citenamefont {Kristinsson}\ \emph {et~al.}(2013)\citenamefont
  {Kristinsson}, \citenamefont {Kyriienko}, \citenamefont {Liew},\ and\
  \citenamefont {Shelykh}}]{kristinsson2013continuous}%
  \BibitemOpen
  \bibfield  {author} {\bibinfo {author} {\bibfnamefont {K.}~\bibnamefont
  {Kristinsson}}, \bibinfo {author} {\bibfnamefont {O.}~\bibnamefont
  {Kyriienko}}, \bibinfo {author} {\bibfnamefont {T.~C.~H.}\ \bibnamefont
  {Liew}},\ and\ \bibinfo {author} {\bibfnamefont {I.~A.}\ \bibnamefont
  {Shelykh}},\ }\bibfield  {title} {\bibinfo {title} {Continuous terahertz
  emission from dipolaritons},\ }\href@noop {} {\bibfield  {journal} {\bibinfo
  {journal} {Physical Review B}\ }\textbf {\bibinfo {volume} {88}},\ \bibinfo
  {pages} {245303} (\bibinfo {year} {2013})}\BibitemShut {NoStop}%
\bibitem [{\citenamefont {Kristinsson}\ \emph {et~al.}(2014)\citenamefont
  {Kristinsson}, \citenamefont {Kyriienko},\ and\ \citenamefont
  {Shelykh}}]{kristinsson2014terahertz}%
  \BibitemOpen
  \bibfield  {author} {\bibinfo {author} {\bibfnamefont {K.}~\bibnamefont
  {Kristinsson}}, \bibinfo {author} {\bibfnamefont {O.}~\bibnamefont
  {Kyriienko}},\ and\ \bibinfo {author} {\bibfnamefont {I.}~\bibnamefont
  {Shelykh}},\ }\bibfield  {title} {\bibinfo {title} {Terahertz laser based on
  dipolaritons},\ }\href@noop {} {\bibfield  {journal} {\bibinfo  {journal}
  {Physical Review A}\ }\textbf {\bibinfo {volume} {89}},\ \bibinfo {pages}
  {023836} (\bibinfo {year} {2014})}\BibitemShut {NoStop}%
\bibitem [{\citenamefont {Sim}\ \emph {et~al.}(2018)\citenamefont {Sim},
  \citenamefont {Lee}, \citenamefont {Trifonov}, \citenamefont {Kim},
  \citenamefont {Cha}, \citenamefont {Sung}, \citenamefont {Cho}, \citenamefont
  {Shim}, \citenamefont {Jo},\ and\ \citenamefont
  {Choi}}]{Sim_Lee_Trifonov_Kim_Cha_Sung_Cho_Shim_Jo_Choi_2018}%
  \BibitemOpen
  \bibfield  {author} {\bibinfo {author} {\bibfnamefont {S.}~\bibnamefont
  {Sim}}, \bibinfo {author} {\bibfnamefont {D.}~\bibnamefont {Lee}}, \bibinfo
  {author} {\bibfnamefont {A.~V.}\ \bibnamefont {Trifonov}}, \bibinfo {author}
  {\bibfnamefont {T.}~\bibnamefont {Kim}}, \bibinfo {author} {\bibfnamefont
  {S.}~\bibnamefont {Cha}}, \bibinfo {author} {\bibfnamefont {J.~H.}\
  \bibnamefont {Sung}}, \bibinfo {author} {\bibfnamefont {S.}~\bibnamefont
  {Cho}}, \bibinfo {author} {\bibfnamefont {W.}~\bibnamefont {Shim}}, \bibinfo
  {author} {\bibfnamefont {M.-H.}\ \bibnamefont {Jo}},\ and\ \bibinfo {author}
  {\bibfnamefont {H.}~\bibnamefont {Choi}},\ }\bibfield  {title} {\bibinfo
  {title} {Ultrafast quantum beats of anisotropic excitons in atomically thin
  res2},\ }\bibfield  {journal} {\bibinfo  {journal} {Nature Communications}\
  }\textbf {\bibinfo {volume} {9}},\ \href
  {https://doi.org/10.1038/s41467-017-02802-8} {10.1038/s41467-017-02802-8}
  (\bibinfo {year} {2018})\BibitemShut {NoStop}%
\bibitem [{\citenamefont {Pei}\ \emph {et~al.}(2017)\citenamefont {Pei},
  \citenamefont {Yang}, \citenamefont {Wang}, \citenamefont {Wang},
  \citenamefont {Mokkapati}, \citenamefont {Lü}, \citenamefont {Zheng},
  \citenamefont {Qin}, \citenamefont {Neshev}, \citenamefont {Tan},\ and\
  \citenamefont {et~al.}}]{Biexcitons}%
  \BibitemOpen
  \bibfield  {author} {\bibinfo {author} {\bibfnamefont {J.}~\bibnamefont
  {Pei}}, \bibinfo {author} {\bibfnamefont {J.}~\bibnamefont {Yang}}, \bibinfo
  {author} {\bibfnamefont {X.}~\bibnamefont {Wang}}, \bibinfo {author}
  {\bibfnamefont {F.}~\bibnamefont {Wang}}, \bibinfo {author} {\bibfnamefont
  {S.}~\bibnamefont {Mokkapati}}, \bibinfo {author} {\bibfnamefont
  {T.}~\bibnamefont {Lü}}, \bibinfo {author} {\bibfnamefont {J.-C.}\
  \bibnamefont {Zheng}}, \bibinfo {author} {\bibfnamefont {Q.}~\bibnamefont
  {Qin}}, \bibinfo {author} {\bibfnamefont {D.}~\bibnamefont {Neshev}},
  \bibinfo {author} {\bibfnamefont {H.~H.}\ \bibnamefont {Tan}},\ and\ \bibinfo
  {author} {\bibnamefont {et~al.}},\ }\bibfield  {title} {\bibinfo {title}
  {Excited state biexcitons in atomically thin mose2},\ }\href
  {https://doi.org/10.1021/acsnano.7b03909} {\bibfield  {journal} {\bibinfo
  {journal} {ACS Nano}\ }\textbf {\bibinfo {volume} {11}},\ \bibinfo {pages}
  {7468–7475} (\bibinfo {year} {2017})}\BibitemShut {NoStop}%
\bibitem [{\citenamefont {Horng}\ \emph {et~al.}(2018)\citenamefont {Horng},
  \citenamefont {Stroucken}, \citenamefont {Zhang}, \citenamefont {Paik},
  \citenamefont {Deng},\ and\ \citenamefont
  {Koch}}]{interlayer_excitons_in_MoSe2_BL}%
  \BibitemOpen
  \bibfield  {author} {\bibinfo {author} {\bibfnamefont {J.}~\bibnamefont
  {Horng}}, \bibinfo {author} {\bibfnamefont {T.}~\bibnamefont {Stroucken}},
  \bibinfo {author} {\bibfnamefont {L.}~\bibnamefont {Zhang}}, \bibinfo
  {author} {\bibfnamefont {E.~Y.}\ \bibnamefont {Paik}}, \bibinfo {author}
  {\bibfnamefont {H.}~\bibnamefont {Deng}},\ and\ \bibinfo {author}
  {\bibfnamefont {S.~W.}\ \bibnamefont {Koch}},\ }\bibfield  {title} {\bibinfo
  {title} {Observation of interlayer excitons in mose2 single crystals},\
  }\bibfield  {journal} {\bibinfo  {journal} {Physical Review B}\ }\textbf
  {\bibinfo {volume} {97}},\ \href {https://doi.org/10.1103/physrevb.97.241404}
  {10.1103/physrevb.97.241404} (\bibinfo {year} {2018})\BibitemShut {NoStop}%
\bibitem [{\citenamefont {Gerber}\ \emph {et~al.}(2019)\citenamefont {Gerber},
  \citenamefont {Courtade}, \citenamefont {Shree}, \citenamefont {Robert},
  \citenamefont {Taniguchi}, \citenamefont {Watanabe}, \citenamefont
  {Balocchi}, \citenamefont {Renucci}, \citenamefont {Lagarde}, \citenamefont
  {Marie},\ and\ \citenamefont
  {et~al.}}]{Gerber_Courtade_Shree_Robert_Taniguchi_Watanabe_Balocchi_Renucci_Lagarde_Marie_2019}%
  \BibitemOpen
  \bibfield  {author} {\bibinfo {author} {\bibfnamefont {I.~C.}\ \bibnamefont
  {Gerber}}, \bibinfo {author} {\bibfnamefont {E.}~\bibnamefont {Courtade}},
  \bibinfo {author} {\bibfnamefont {S.}~\bibnamefont {Shree}}, \bibinfo
  {author} {\bibfnamefont {C.}~\bibnamefont {Robert}}, \bibinfo {author}
  {\bibfnamefont {T.}~\bibnamefont {Taniguchi}}, \bibinfo {author}
  {\bibfnamefont {K.}~\bibnamefont {Watanabe}}, \bibinfo {author}
  {\bibfnamefont {A.}~\bibnamefont {Balocchi}}, \bibinfo {author}
  {\bibfnamefont {P.}~\bibnamefont {Renucci}}, \bibinfo {author} {\bibfnamefont
  {D.}~\bibnamefont {Lagarde}}, \bibinfo {author} {\bibfnamefont
  {X.}~\bibnamefont {Marie}},\ and\ \bibinfo {author} {\bibnamefont {et~al.}},\
  }\bibfield  {title} {\bibinfo {title} {Interlayer excitons in bilayer mos2
  with strong oscillator strength up to room temperature},\ }\bibfield
  {journal} {\bibinfo  {journal} {Physical Review B}\ }\textbf {\bibinfo
  {volume} {99}},\ \href {https://doi.org/10.1103/physrevb.99.035443}
  {10.1103/physrevb.99.035443} (\bibinfo {year} {2019})\BibitemShut {NoStop}%
\bibitem [{\citenamefont {Schuller}\ \emph {et~al.}(2013)\citenamefont
  {Schuller}, \citenamefont {Karaveli}, \citenamefont {Schiros}, \citenamefont
  {He}, \citenamefont {Yang}, \citenamefont {Kymissis}, \citenamefont {Shan},\
  and\ \citenamefont {Zia}}]{schuller2013orientation}%
  \BibitemOpen
  \bibfield  {author} {\bibinfo {author} {\bibfnamefont {J.~A.}\ \bibnamefont
  {Schuller}}, \bibinfo {author} {\bibfnamefont {S.}~\bibnamefont {Karaveli}},
  \bibinfo {author} {\bibfnamefont {T.}~\bibnamefont {Schiros}}, \bibinfo
  {author} {\bibfnamefont {K.}~\bibnamefont {He}}, \bibinfo {author}
  {\bibfnamefont {S.}~\bibnamefont {Yang}}, \bibinfo {author} {\bibfnamefont
  {I.}~\bibnamefont {Kymissis}}, \bibinfo {author} {\bibfnamefont
  {J.}~\bibnamefont {Shan}},\ and\ \bibinfo {author} {\bibfnamefont
  {R.}~\bibnamefont {Zia}},\ }\bibfield  {title} {\bibinfo {title} {Orientation
  of luminescent excitons in layered nanomaterials},\ }\href@noop {} {\bibfield
   {journal} {\bibinfo  {journal} {Nature nanotechnology}\ }\textbf {\bibinfo
  {volume} {8}},\ \bibinfo {pages} {271} (\bibinfo {year} {2013})}\BibitemShut
  {NoStop}%
\bibitem [{\citenamefont {Wang}\ \emph {et~al.}(2017)\citenamefont {Wang},
  \citenamefont {Chiu}, \citenamefont {Honz}, \citenamefont {Mak},\ and\
  \citenamefont {Shan}}]{wang2017electrical}%
  \BibitemOpen
  \bibfield  {author} {\bibinfo {author} {\bibfnamefont {Z.}~\bibnamefont
  {Wang}}, \bibinfo {author} {\bibfnamefont {Y.-H.}\ \bibnamefont {Chiu}},
  \bibinfo {author} {\bibfnamefont {K.}~\bibnamefont {Honz}}, \bibinfo {author}
  {\bibfnamefont {K.~F.}\ \bibnamefont {Mak}},\ and\ \bibinfo {author}
  {\bibfnamefont {J.}~\bibnamefont {Shan}},\ }\bibfield  {title} {\bibinfo
  {title} {Electrical tuning of interlayer exciton gases in wse2 bilayers},\
  }\href@noop {} {\bibfield  {journal} {\bibinfo  {journal} {Nano letters}\
  }\textbf {\bibinfo {volume} {18}},\ \bibinfo {pages} {137} (\bibinfo {year}
  {2017})}\BibitemShut {NoStop}%
\bibitem [{\citenamefont {Jakubczyk}\ \emph {et~al.}(2016)\citenamefont
  {Jakubczyk}, \citenamefont {Delmonte}, \citenamefont {Koperski},
  \citenamefont {Nogajewski}, \citenamefont {Faugeras}, \citenamefont
  {Langbein}, \citenamefont {Potemski},\ and\ \citenamefont
  {Kasprzak}}]{FWM_Tr_NX}%
  \BibitemOpen
  \bibfield  {author} {\bibinfo {author} {\bibfnamefont {T.}~\bibnamefont
  {Jakubczyk}}, \bibinfo {author} {\bibfnamefont {V.}~\bibnamefont {Delmonte}},
  \bibinfo {author} {\bibfnamefont {M.}~\bibnamefont {Koperski}}, \bibinfo
  {author} {\bibfnamefont {K.}~\bibnamefont {Nogajewski}}, \bibinfo {author}
  {\bibfnamefont {C.}~\bibnamefont {Faugeras}}, \bibinfo {author}
  {\bibfnamefont {W.}~\bibnamefont {Langbein}}, \bibinfo {author}
  {\bibfnamefont {M.}~\bibnamefont {Potemski}},\ and\ \bibinfo {author}
  {\bibfnamefont {J.}~\bibnamefont {Kasprzak}},\ }\bibfield  {title} {\bibinfo
  {title} {Radiatively limited dephasing and exciton dynamics in mose$_2$
  monolayers revealed with four-wave mixing microscopy},\ }\href
  {https://doi.org/10.1021/acs.nanolett.6b01060} {\bibfield  {journal}
  {\bibinfo  {journal} {Nano Letters}\ }\textbf {\bibinfo {volume} {16}},\
  \bibinfo {pages} {5333–5339} (\bibinfo {year} {2016})}\BibitemShut
  {NoStop}%
\bibitem [{\citenamefont {Shepard}\ \emph {et~al.}(2017)\citenamefont
  {Shepard}, \citenamefont {Ardelean}, \citenamefont {Ajayi}, \citenamefont
  {Rhodes}, \citenamefont {Zhu}, \citenamefont {Hone},\ and\ \citenamefont
  {Strauf}}]{FWM_Tr}%
  \BibitemOpen
  \bibfield  {author} {\bibinfo {author} {\bibfnamefont {G.~D.}\ \bibnamefont
  {Shepard}}, \bibinfo {author} {\bibfnamefont {J.~V.}\ \bibnamefont
  {Ardelean}}, \bibinfo {author} {\bibfnamefont {O.~A.}\ \bibnamefont {Ajayi}},
  \bibinfo {author} {\bibfnamefont {D.}~\bibnamefont {Rhodes}}, \bibinfo
  {author} {\bibfnamefont {X.}~\bibnamefont {Zhu}}, \bibinfo {author}
  {\bibfnamefont {J.~C.}\ \bibnamefont {Hone}},\ and\ \bibinfo {author}
  {\bibfnamefont {S.}~\bibnamefont {Strauf}},\ }\bibfield  {title} {\bibinfo
  {title} {Trion-species-resolved quantum beats in mose$_2$},\ }\href
  {https://doi.org/10.1021/acsnano.7b06444} {\bibfield  {journal} {\bibinfo
  {journal} {ACS Nano}\ }\textbf {\bibinfo {volume} {11}},\ \bibinfo {pages}
  {11550–11558} (\bibinfo {year} {2017})}\BibitemShut {NoStop}%
\bibitem [{\citenamefont {Moody}\ \emph {et~al.}(2015)\citenamefont {Moody},
  \citenamefont {Kavir~Dass}, \citenamefont {Hao}, \citenamefont {Chen},
  \citenamefont {Li}, \citenamefont {Singh}, \citenamefont {Tran},
  \citenamefont {Clark}, \citenamefont {Xu}, \citenamefont {Berghäuser},\ and\
  \citenamefont {et~al.}}]{fwm_NX}%
  \BibitemOpen
  \bibfield  {author} {\bibinfo {author} {\bibfnamefont {G.}~\bibnamefont
  {Moody}}, \bibinfo {author} {\bibfnamefont {C.}~\bibnamefont {Kavir~Dass}},
  \bibinfo {author} {\bibfnamefont {K.}~\bibnamefont {Hao}}, \bibinfo {author}
  {\bibfnamefont {C.-H.}\ \bibnamefont {Chen}}, \bibinfo {author}
  {\bibfnamefont {L.-J.}\ \bibnamefont {Li}}, \bibinfo {author} {\bibfnamefont
  {A.}~\bibnamefont {Singh}}, \bibinfo {author} {\bibfnamefont
  {K.}~\bibnamefont {Tran}}, \bibinfo {author} {\bibfnamefont {G.}~\bibnamefont
  {Clark}}, \bibinfo {author} {\bibfnamefont {X.}~\bibnamefont {Xu}}, \bibinfo
  {author} {\bibfnamefont {G.}~\bibnamefont {Berghäuser}},\ and\ \bibinfo
  {author} {\bibnamefont {et~al.}},\ }\bibfield  {title} {\bibinfo {title}
  {Intrinsic homogeneous linewidth and broadening mechanisms of excitons in
  monolayer transition metal dichalcogenides},\ }\bibfield  {journal} {\bibinfo
   {journal} {Nature Communications}\ }\textbf {\bibinfo {volume} {6}},\ \href
  {https://doi.org/10.1038/ncomms9315} {10.1038/ncomms9315} (\bibinfo {year}
  {2015})\BibitemShut {NoStop}%
\end{thebibliography}%

\end{document}